\documentclass[a4paper,11pt]{article}
\pdfoutput=1

\usepackage{jheppub} 

\usepackage[T1]{fontenc} 

\usepackage{amssymb}\usepackage[normalem]{ulem} % delete me
\usepackage{ulem}

\usepackage{verbatim}
\usepackage{graphicx} % Required for inserting images
\usepackage{graphics}
\usepackage{epsfig} 
\usepackage{color}
\usepackage{amssymb}
\usepackage{amsmath}
\usepackage{doi}
\usepackage{dsfont}
\usepackage{setspace}
\usepackage{float}
\usepackage{caption}
\usepackage{subcaption}

\usepackage{slashed}
\usepackage{booktabs}
\usepackage{multirow}
\usepackage{soul}

\title{NLO QCD predictions for $\boldsymbol{t\bar{t}\gamma}$ with realistic photon isolation}

\author[\, a, b]{Daniel Stremmer}
\author[a]{and Malgorzata Worek\,}

\affiliation[a]{Institute for Theoretical Particle Physics
and Cosmology, RWTH Aachen University, \\D-52056 Aachen, Germany}
\affiliation[b]{Institute for Theoretical Particle Physics,
Karlsruhe Institute of Technology, \\D-76128 Karlsruhe, Germany}

\emailAdd{daniel.stremmer@kit.edu}
\emailAdd{worek@physik.rwth-aachen.de}

\abstract{We present a complete description of top quark pair production in association with a hard photon in the di-lepton decay channel. The calculation is performed at  NLO QCD and includes all resonant and non-resonant Feynman diagrams, interferences, and finite-width effects of the top quarks and $W^\pm/Z$ gauge bosons. We provide the results for the $pp\to e^+\nu_e \,\mu^- \bar{\nu}\, b\bar{b}\,\gamma+X$ process using the fixed-cone, smooth-cone and hybrid-photon isolation criteria.  The fixed-cone isolation criterion allows contributions from collinear photon radiation off QCD partons, which requires the inclusion of parton-to-photon fragmention processes. To this end, we include the latter contributions  into our computational framework. We quantify the impact of different photon-isolation prescriptions on the integrated and differential cross-section predictions for the LHC at a centre-of-mass energy of $\sqrt{s}=13.6$ TeV. Our state-of-the-art NLO QCD results with the fixed-cone criterion allow us to reproduce the photon-isolation prescription employed in  ATLAS and CMS. This will help to improve future comparisons with the LHC data.}

\dedicated{\rm TTK-24-31, P3H-24-063}


\begin{document} 

\maketitle
\flushbottom

%-----------------------------------------------------------
%
\section{Introduction}
%
%-----------------------------------------------------------

Processes with photons, which occur in $pp$ collisions at the Large Hadron Collider (LHC), play a central role in the ATLAS and CMS physics programs. Such processes are dominated by all phenomena in which photons come from hadron decays, such as for example neutral pion $(\pi^0\to \gamma\gamma)$ or eta meson decays $(\eta\to \gamma\gamma)$. However, dedicated experimental measurements at the LHC are mostly interested in primary (also called direct or prompt) photons. Such photons appear much less frequently. They are produced in the hard scattering process before the quarks and gluons have time to form hadrons and long before these hadrons have time to decay. The production of prompt photons ranges from associated production processes, where photons are produced in combination with heavy quarks, jets, massive bosons and other photon(s) to processes in which  photons arise in the decay of heavy resonances. Physics analyses and measurements at the LHC involving photons test the perturbative and non-perturbative regimes of QCD and the electroweak sector of the Standard Model (SM). Moreover, they are used to search for signals of new physics beyond the SM (BSM). Photons are also produced in the transition to hadrons which can be described by fragmentation functions. These  fragmentation functions are conceptually similar to parton distribution functions but describe the probability for the transition of a parton into a collinear photon with a given momentum fraction.  As one might expect,  identifying photons in hadron collisions is a  difficult task as many other particles are created near them in the detector. Because direct photons are rather rare phenomena, very good photon identification capabilities and precise isolation procedures are essential for the extensive physics program currently underway at the LHC. Such procedures will become even more important in the near future with the planned High Luminosity phase of the LHC (HL-LHC), which is expected to significantly increase the discovery potential of ATLAS and  CMS by increasing the integrated luminosity up to  even $(3000 - 4000)~{\rm  fb}{}^{-1}$  \cite{ZurbanoFernandez:2020cco}. At the LHC, the photon candidates are reconstructed from energy deposits  in the central region of the ATLAS/CMS electromagnetic calorimeter and by requiring them to be isolated from other particles in the event. The isolation criterion is usually formulated by allowing only a limited amount of hadronic energy in a fixed cone of size $R$ around the direction of the photon. Such an isolation prescription is very effective in suppressing  photons emerging in decays of strongly interacting particles, but it allows for contributions from photon fragmentation processes. Therefore, theoretical predictions for processes with prompt photons must take into account both direct and fragmentation contributions. Only the combined result can be subjected to the fixed-cone isolation prescription for infrared-safe observables. The way out of this situation is to impose the smooth-cone isolation prescription  \cite{Frixione:1998jh}. This approach eliminates fragmentation contributions by smoothly lowering the amount of hadronic energy towards the center of the cone of size $R$. In the strict collinear limit, when $R \rightarrow 0$, any emission will lead to the photon being considered as not isolated. The latter condition is mandatory for higher-order (perturbative) calculations because the phase space for soft radiation cannot be modified to ensure the adequate cancellation of infrared divergences between the real emission and the virtual contribution. This theoretically very well-founded and appealing method of isolating photons is used in almost all high-precision theoretical predictions with prompt photons at the LHC. However, such a method cannot be directly applied to experimental measurements. Although this procedure has several arbitrary parameters that can in principle be tuned to reflect what happens on the experimental side, such tuning depends greatly on the process under consideration and the regions of phase space being analysed. As a result, it is rarely performed. The smooth-cone isolation prescription therefore introduces an additional systematic uncertainty in comparisons between theoretical results and LHC data. An alternative isolation prescription, called the hybrid-photon isolation method, has been employed in Ref. \cite{Siegert:2016bre,Chen:2019zmr}. In this approach the smooth-cone isolation prescription with a small radius $R$ is combined with the fixed-cone isolation method for $R_{\rm fixed}$, such that  $R_{\rm fixed} \gg R$. The smooth-cone prescription, that is applied in the first step, removes collinear divergences and the dependence on the fragmentation processes. On the other hand, the fixed-cone isolation, that is applied in the next step, makes a comparison to the LHC data straightforward. Also in this case the cone size $R$ can be tuned to reproduce the behaviour of the fixed-cone isolation approach that is used in both ATLAS and CMS. Nevertheless, choosing  $R$ too small might generate large logarithms of the form $\ln(R)$ that must be properly resummed. These three approaches appear to be quite different conceptually. Therefore, it would be beneficial to thoroughly investigate their advantages and disadvantages in the case of a realistic, multiparticle final state that includes not only photon emission in production and decays of  heavy resonances but also additional jet activity. 

The aim of this work is to make such a comparison at the NLO QCD level for the $pp \to e^+ \nu_e \, \mu^- \bar{\nu}_\mu \, b\bar{b} \, \gamma+X$ process.  In our calculation, photons are included in the production phase of the process as well as in top-quark decays. Furthermore, not only double-resonant top-quark and $W$ gauge boson contributions but also single-resonant and non-resonant contributions are taken into account. In addition,  the Breit-Wigner propagators are incorporated for heavy resonances appearing in the process. To perform  NLO QCD studies with the fixed-cone isolation, two different parton-to-photon fragmentation functions are build in into our computational framework. The second goal of the paper is to provide the state-of-the art NLO QCD predictions for $pp \to e^+ \nu_e \, \mu^- \bar{\nu}_\mu \, b\bar{b} \, \gamma+X$ using the LHC Run III center-of-mass energy of  $\sqrt{s}=13.6$ TeV. As is customary in this type of studies we will provide the integrated and differential cross-section results for different renormalisation and factorisation scale settings. Thanks to current higher-order calculations for this process, future comparisons with  the LHC data will become more accurate and precise.

Let us mention here that on the theory side,  NLO QCD and NLO electroweak (EW) corrections to the $pp \to t\bar{t}\gamma+X$  process have been calculated long time ago \cite{Duan:2009kcr,Duan:2011zxb,Maltoni:2015ena,Duan:2016qlc}, albeit only  for stable top quarks. Recently, also the so-called complete NLO predictions for this process have been provided \cite{Pagani:2021iwa}. These complete theoretical predictions comprise all leading and subleading LO contributions as well as their corresponding higher-order QCD and EW effects. The approximate NNLO cross section, with second-order soft-gluon corrections added to the NLO result including QCD and EW higher-order contributions, has also been calculated \cite{Kidonakis:2022qvz}. In addition,  various predictions that take into account top-quark decays, are available in the literature. First, NLO QCD theoretical predictions for  $pp \to t\bar{t}\gamma+X$ have been matched with the \textsc{Pythia} parton shower program \cite{Kardos:2014zba}. Theoretical predictions at NLO in QCD in the Narrow Width Approximation (NWA) have been provided  \cite{Melnikov:2011ta,Bevilacqua:2019quz}. A fully realistic description of this process at NLO in QCD is also available in the literature \cite{Bevilacqua:2018woc,Bevilacqua:2018dny}. This calculation is based on matrix elements for the $e^+\nu_e \, \mu^- \bar{\nu}_\mu \, b\bar{b}\, \gamma$ final state and allows the production of $t\bar{t}\gamma$, $tW\gamma$, $W^+W^-b\bar{b}\gamma$ events including interference and off-shell effects of top quarks and $W^\pm/Z$ gauge bosons. Moreover, a dedicated comparison between the full off-shell calculation and the  result in the NWA has also been carried out \cite{Bevilacqua:2019quz}. Finally, the complete NLO corrections for the $pp \to t\bar{t}\gamma+X$ process have recently been calculated in the di-lepton decay channel   \cite{Stremmer:2024ecl}. In this study, the NWA  has been employed to analyse the individual size of each subleading contribution and the origin of the  leading and subleading QCD and EW corrections.

The rest of the paper is organised as follows. In Section \ref{isolated-photons} we give definitions of the various photon-isolation prescriptions. The theoretical framework for our calculations is summarised in Section \ref{theoretical-framework}, where we also provide the explicit expressions for the LO and NLO QCD cross sections. In Section \ref{computational-setup} we describe our computational framework, together with modifications made in the \textsc{Helac-Dipoles} Monte Carlo program to include the photon fragmentation function. Our theoretical setup is described in Section \ref{input-parameteres}. In Section  \ref{comparison-integrated} and Section \ref{comparison-differential}  we compare the results obtained using the fixed-cone, smooth-cone and hybrid-photon isolation prescription at the integrated and differential cross-section level. The state-of-the-art NLO QCD predictions comprising full off-shell effects and the fixed-cone isolation are given in Section \ref{results-fixed-cone}. Our results are summarised in Section \ref{summary}. 

%-----------------------------------------------------------
%
\section{How to isolate a prompt photon}
\label{isolated-photons}
%
%-----------------------------------------------------------

To separate prompt photons produced by hard scattering processes from photons from other sources, an isolation requirement must be imposed.  There are several examples of such isolation prescriptions. Below we briefly describe the following three approaches, that will be employed in our studies:
\begin{itemize}
\item
fixed-cone isolation,
\item 
smooth-cone isolation,
\item 
hybrid-photon isolation.
\end{itemize}
All of which are based on the use of some kind of cone around the photon candidate.

%-----------------------------------------------------------
%
\subsection*{Fixed-cone isolation}
%
%-----------------------------------------------------------

The most straightforward way to identify a photon is to first define a fixed cone  of angular size $R$ around a photon candidate.  Because we want the photon to be isolated, it must carry the largest part of the (transverse) energy in its vicinity. Furthermore, the photon is required to be isolated from other objects within this cone by demanding that the hadronic transverse energy of the final-state particles within the cone of size $R$, labelled here as  $E_T^{\rm had}$, is smaller than a certain value given by  
\begin{equation}
\label{fixed-cone}
E_T^{\rm had}(R) \le E_{T}^{\rm max}(p_{T,\,\gamma}) = \varepsilon_\gamma\, p_{T,\,\gamma} + E_T^{\rm thres}\,,
\end{equation}
where the free parameters: $R, \, \varepsilon_\gamma$, and $E_T^{\rm thres}$ are chosen for a particular process with photons, while $p_{T,\,\gamma}$ is the transverse momentum of the photon. The fixed-cone isolation prescription is commonly used in various measurements by ATLAS and CMS. With this isolation criterion the cross-section computation needs to include the non-perturbative photon fragmentation functions to be collinear finite.  Such fragmentation functions are poorly known and may introduce additional uncertainties, the magnitude of which is not  known in advance. We note here, that taking the limit of $E_{T}^{\rm max}(p_{T,\,\gamma}) \rightarrow 0$ eliminates any fragmentation contribution but renders the cross sections infrared unsafe.

%-----------------------------------------------------------
%
\subsection*{Smooth-cone isolation}
%
%-----------------------------------------------------------

Another way of defining the isolation criterion has been introduced in Ref.  \cite{Frixione:1998jh}. In this case the isolation condition depends directly on $R$. In this approach a cone of fixed radius $R_0$ is first drawn around the photon candidate. Then for all $R \le R_0$  the total amount of hadronic  transverse energy inside the $R$ cone is required to satisfy the following condition
\begin{equation}
E_T^{\rm had}(R) \le E_{T}^{\rm max}(p_{T,\,\gamma} \,, R) \,,
\end{equation}
where the energy profile $E_{T}^{\rm max}(p_{T,\,\gamma} \,, R)$ is largely arbitrary. However, it  must be some continuous function of $R$ that satisfies the following conditions: it increases with $R$ and decreases to zero when $R \rightarrow 0$. The following form has been proposed in  Ref. \cite{Frixione:1998jh}
\begin{equation}
E_T^{\rm had}(R) = \sum_{i} p_{T,\, i}\, \Theta (R-R_{\gamma i}) \le E_{T}^{\rm max}(p_{T,\,\gamma} \,, R) = \varepsilon_\gamma \,  p_{T, \, \gamma} \left( \frac{1 -\cos R}{1 -\cos R_0}\right)^n\,,
\end{equation}
where $R_0, \, \varepsilon_\gamma$, and $n$ are the  parameters with $\varepsilon_\gamma$ and $n$ being positive numbers of the order of $1$. In addition,  $R_{\gamma i}$ is defined according to 
\begin{equation}
R_{\gamma i} =\sqrt{(y_\gamma - y_i)^2 + (\phi_\gamma - \phi_i)^2}\,.
\end{equation}
The sum $\sum_i$ runs over all partons,  $p_{T,\, i}$ is the transverse momentum of the parton and $p_{T, \, \gamma}$ is the transverse momentum of the photon. This isolation criterion allows arbitrarily soft radiation inside the cone of size $R$, but collinear  radiation $(\cos R  \rightarrow 1)$ is forbidden. Thus, the contributions from the fragmentation functions are simply eliminated. Having infrared finite cross sections without the need to subtract collinear QED final-state singularities is a significant technical simplification. As a result, this approach has been successfully applied to obtain higher-order theoretical predictions for processes with photons and jets both at the NLO and NNLO level in QCD. On the other hand, the smooth nature of this isolation prescription cannot be easily reproduced in experimental measurements at the LHC where detectors with finite granularity are employed.  In consequence, the smooth-cone isolation criterion introduces a systematic uncertainty in comparisons with the LHC data. The latter uncertainty  can be significantly reduced by tuning the free parameters $\varepsilon_\gamma$, and $n$ of the smooth-cone isolation prescription to mimic the effects of the fixed-cone isolation approach. However, such a procedure is time-consuming and not very practical, since it must be repeated for each process considered and set of phase-space cuts applied. We would like to note here that from the theory point of view it would be enough, at least at NLO, to impose an angular separation between quark-initiated jets and photons to remove collinear infrared singularities and consequently fragmentation contributions. However, at the LHC, jets initiated by quarks and gluons cannot be well distinguished. Therefore, in order to closely follow the experimental setup the same event selection has to be applied to all types of jets. Setting simply $\Delta R> 0.4$  between photons and jets would introduce an angular separation between photons and gluons which would spoil the cancellation of infrared singularities between the real and virtual contributions in higher-order calculations. 

%-----------------------------------------------------------
%
\subsection*{Hybrid-photon isolation}
%
%-----------------------------------------------------------

In order to reduce the dependence on the arbitrary input parameters in the smooth-cone isolation condition and simultaneously improve the agreement with the fixed-cone isolation condition, the hybrid-photon isolation, as employed in Ref. \cite{Siegert:2016bre,Chen:2019zmr}, can be used. In this approach, the smooth-cone and fixed-cone isolation methods are combined together. In the first step a smaller cone with radius $R$ is defined within the cone size $R_{\rm fixed}$, such that $R < R_{\rm fixed}$. After removing the fragmentation contributions inside the inner cone $R$ in an infrared-safe manner with the help of  the smooth-cone isolation method, the outer cone of size $R_{\rm fixed}$ is used to apply the experimental fixed-cone isolation with the exact parameters as in the experimental setups. Thus, it is expected that the differences with respect to the fixed-cone isolation decrease and that the dependence on the input parameters of the (inner) smooth-cone isolation is drastically reduced. In fact, the inner smooth-photon isolation procedure rejects only a small number of events if the size of $R$ is chosen small enough. However,  for very small radius $R$ the cross-section predictions might in principle become very large, indicating a breakdown of the fixed-order perturbation theory. In such a case, the leading $\ln (R)$ terms have to be resummed, see e.g. Refs. \cite{Catani:2002ny,Catani:2013oma,Becher:2022rhu}.

In what follows, we will conduct a comparison of the three isolation criteria  and examine the effect of tuning the relevant parameters on the  integrated and differential  cross-section predictions for the  $pp\to e^+\nu_e \, \mu^- \bar{\nu}_\mu \, b\bar{b} \, \gamma+X$ process at the LHC with $\sqrt{s}=13.6$ TeV. Taking into account all the advantages and disadvantages of the described methods, it is important to analyse them in more detail within a common computational framework. 

%-----------------------------------------------------------
%
\section{Description of the calculation}
\label{theoretical-framework}
%
%-----------------------------------------------------------

At higher orders in QCD, any  cross section for a process involving photons and jets consists of a direct and a fragmentation contribution. In particular,  for the $pp \to e^+\nu_e \, \mu^- \bar{\nu}_\mu  \,b\bar{b}\,\gamma+X$ process (denoted here as  $pp\to t\bar{t}\gamma+X$) when following the notation of Ref. \cite{Gehrmann:2022cih} we can write it  as
\begin{equation}
\label{general}
 d\hat{\sigma}_{t\bar{t}\gamma+X}= d\hat{\sigma}_{t\bar{t}\gamma} + \sum_{p} d\hat{\sigma}_{t\bar{t}p} \otimes  D^{B}_{p\to \gamma}\,,
\quad \quad \quad \quad \quad \quad p\in\left\{q_j,\bar{q}_j,g\right\}\,.
\end{equation}
The first contribution corresponds to the prompt-photon production, where the photon is produced directly in the hard interaction. The second term comprises the longer distance fragmentation process with one of the final-state partons fragmenting into a photon and transferring a fraction of its momentum to the photon. Apart from the leading order level, the direct $d\hat{\sigma}_{t\bar{t}\gamma}$ cross section contains singularities originating from configurations in which the massless final-state partons are collinear with the photon. Because of the general and process-independent structure of collinear divergences, it is possible to absorb the uncancelled divergences into the fragmentation functions. Even though the fragmentation functions are non-perturbative, we can assign powers of  coupling constants to them by counting the couplings necessary to emit a photon. Consequently, $D_{q \to \gamma}$ is of the order of ${\cal O}(\alpha)$. Since the gluon can only couple to the photon via a quark $D_{g\to \gamma}$  is of the order of ${\cal O}(\alpha \alpha_s)$. The fragmentation of gluons into photons is therefore a much smaller effect than the fragmentation of quarks into photons. In Eq. \eqref{general} the bare fragmentation contribution $D^{B}_{p\to \gamma}$ can be further decomposed into a piece where $d\hat{\sigma}_{t\bar{t}p}$ is convoluted with the renormalised fragmentation functions and the collinear  counterterms of the fragmentation functions. The later contribution is responsible for the cancellation of  the parton-photon collinear singularities in the direct contribution. This factorisation is performed at a fragmentation scale $\mu_{Frag}$, and the resulting  fragmentation functions consequently depend on this scale choice. The relation between renormalised  and bare fragmentation functions can be expressed as
\begin{equation}
\label{eq_3.2}
D_{i \to \gamma}(z,\mu^2_{Frag}) =\sum_{j} \mathbf{\Gamma}_{i\to j} (z,\mu^2_{Frag})  \otimes D_{j\to \gamma}^{B}(z)\,,
\end{equation}
where $z$ is the  photon momentum fraction, $i,j\in\left\{ q,\bar{q},g,\gamma\right\}$ and $\mathbf{\Gamma}_{i\to j}$ are the factorisation kernels of the fragmentation functions that carry colour factors. Moreover, in order to write this expression in a compact form using $\sum_j$, a photon-to-photon fragmentation function of the following form has to be  introduced
\begin{equation}
D_{\gamma\to \gamma} (z,\mu^2_{Frag}) = D^B_{\gamma\to \gamma} (z) = \delta (1-z)\,.
\end{equation}
The perturbative expansion of the bare fragmentation function is obtained by inverting Eq.~\eqref{eq_3.2} and then using the fact that the factorisation kernels $\mathbf{\Gamma}_{i \to j}$ (or equivalently their inverse) can be expanded perturbatively in $\alpha$ and $\alpha_s$.  For the quark-to-photon fragmentation function relevant for NLO calculations we obtain 
\begin{equation}
D^{B}_{q\to \gamma}(z) = D_{q\to \gamma} (z,\mu^2_{Frag}) - \frac{\alpha}{2\pi} \mathbf{\Gamma}_{q\to \gamma}^{(0)}\,,
\end{equation}
where the subscript $(0)$ indicates the leading term in the expansion of the factorisation kernel.  A similar expression can be provided for the anti-quark-to-photon fragmentation functions.  With the concrete power counting we have chosen to assign to the fragmentation functions, i.e. $D_{q\to\gamma} \sim {\cal O}(\alpha)$ and $D_{g\to \gamma} \sim {\cal O}(\alpha \alpha_s)$, it is clear that there cannot be any contributions from the gluon-to-photon fragmentation function at NLO in QCD. However, they become mandatory at higher orders in $\alpha_s$. We would like to emphasise at this point that if we used a different convention with a different power counting, the $D_{g\to \gamma}$ contribution could already exist at the NLO level in QCD, see e.g. Ref. \cite{Chen:2022gpk}. With our convention for the power counting of the fragmentation functions, the different contributions to the cross section with the photon can be finally written as 
\begin{equation}
\begin{split}
d\hat{\sigma}_{t\bar{t}\gamma+X}^{{\rm  LO}} & = d\hat{\sigma}_{t\bar{t}\gamma+X}^{ {\rm  LO}} \,,\\
d\hat{\sigma}_{t\bar{t}\gamma+X}^{{\rm  NLO}} & = d\hat{\sigma}_{t\bar{t}\gamma+X}^{ {\rm  NLO}} 
+ \sum_q d\hat{\sigma}_{q}^{{\rm  LO}} \otimes  D_{q \to \gamma} - \sum_q d\hat{\sigma}_{q}^{ {\rm  LO}} \otimes \frac{\alpha}{2\pi} \mathbf{\Gamma}^{(0)}_{q \to \gamma}\,,
 \end{split}
\end{equation}
where the sums cover all active quark and anti-quark flavours. Alternatively, we can rewrite this cross section as
\begin{equation}
d\hat{\sigma}_{t\bar{t}\gamma+X}^{\rm NLO}
=\left(d\hat{\sigma}_{t\bar{t}\gamma+X}^{\rm NLO}\right)_{\rm dir}
+\left(d\hat{\sigma}_{t\bar{t}\gamma+X}^{\rm NLO}\right)_{\rm frag}\,,
\end{equation}
where the full calculation is given by the sum of the direct and the fragmentation contribution. They are both IR finite and are given by
\begin{equation} \label{eq:cross_pho_dir}
\left(d\hat{\sigma}_{t\bar{t}\gamma+X}^{\rm NLO}\right)_{\rm dir}
=d\hat{\sigma}_{t\bar{t}\gamma+X}^{\rm NLO}-\frac{\alpha}{2\pi}\sum_{p} d\hat{\sigma}_{p}^{\rm LO} \otimes \mathbf{\Gamma}^{(0)}_{p\to \gamma} \,,
\end{equation}
and
\begin{equation} \label{eq:cross_pho_frag}
\left(d\hat{\sigma}_{t\bar{t}\gamma+X}^{\rm NLO} \right)_{\rm frag}
=\sum_{p} d\hat{\sigma}_{p}^{\rm LO} \otimes D_{p\to \gamma} \,.
\end{equation}
Within the fixed-order approach, only the prompt photon process contributes at lowest order.  At NLO in QCD both prompt photon production and the quark-to-photon fragmentation process contribute. Prompt photon production may occur via the one loop virtual corrections  or by emission of an additional parton to the Born-level process. The fragmentation contribution arises from the latter part with one quark fragmenting into a photon.  The remaining unfragmented partons are subjected to a jet clustering algorithm to finally identify them as jets.  Collinear singularities that  occur when the quark and photon become collinear are factorisable and can be absorbed by a redefinition of the fragmentation functions as described above. The factorisation kernel of the quark-to-photon fragmentation functions can be obtained e.g. from Refs. \cite{Gehrmann-DeRidder:1997fom,Gehrmann:2022cih}
\begin{equation} \label{eq:frag_ct}
\mathbf{\Gamma}^{(0)}_{q\to \gamma}= Q_q^2\, \frac{(4\pi)^{\epsilon}}{\Gamma(1-\epsilon)}\left(\frac{\mu_R^2}{\mu_{Frag}^2}\right)^{\epsilon}\left( - \frac{1}{\epsilon} P_{q\to\gamma}(z) \right)\,,
\end{equation}
where the quark-to-photon splitting function $P_{q\to\gamma}(z)$ is given by
\begin{equation}
P_{q\to\gamma}(z)=\frac{1+(1-z)^2}{z}\,.
\end{equation}
The introduced scale dependence into the renormalised fragmentation functions  is described by the (time-like) Dokshitzer-Gribov-Lipatov-Altarelli-Parisi (DGLAP) evolution equations. Similar to the case of PDFs, evolution equations can be derived for the fragmentation functions by requiring that the bare fragmentation functions are independent of the fragmentation scale
\begin{equation}
    \mu_{Frag}^2\,\frac{d D_{p\to\gamma}^{B}(z)}{d \mu_{Frag}^2}=0.
\end{equation}
This equation can be expanded up to $\mathcal{O}(\alpha)$, which then leads to the following evolution equation for the quark-to-photon fragmentation functions
\begin{equation}
    \mu_{Frag}^2\,\frac{\partial D_{q\to\gamma}(z,\mu_{Frag})}{\partial \mu_{Frag}^2}=\left(\frac{\alpha Q_q^2}{2\pi}\right)P_{q\to\gamma}(z)\,.
\end{equation}
The later equation completely determines the $\mu_{Frag}$ dependence of the LO fragmentation functions. The solution of $D_{q\to\gamma}$ at LO is then given by
\begin{equation} \label{eq:frag_aleph_lo}
    D_{q\to\gamma}^{\rm (LO)}(z,\mu_{Frag})=D_{q\to\gamma}^{np}(z,\mu_0)+\left(\frac{\alpha Q_q^2}{2\pi}\right)P_{q\to\gamma}(z)\,\log\left(\frac{\mu_{Frag}^2}{\mu_0^2}\right),
\end{equation}
where $D_{q\to\gamma}^{np}(z,\mu_0^2)$ at the scale $\mu_0^2$ is a non-perturbative input that  must be obtained by comparisons with experimental measurements. 

A first determination of the LO quark-to-photon fragmentation function has been performed by the  ALEPH collaboration \cite{ALEPH:1995zdi}.   The NLO quark-to-photon fragmentation function has been provided in Refs. \cite{Gehrmann-DeRidder:1997fhc,Gehrmann-DeRidder:1997fom}. The non-perturbative input, $D_{q\to\gamma}^{np}(z,\mu_0^2)$, has been obtained by a fit to the $e^+e^-\to \gamma+ 1j$ data from ALEPH  \cite{ALEPH:1995zdi}. Its explicit form obtained from the lowest-order fit can be found in e.g. Refs. \cite{ALEPH:1995zdi,Gehrmann-DeRidder:1997fom}.  The fragmentation functions of quarks and gluons into photons  with Beyond Leading Logarithm corrections (BLL) have been provided in Ref. \cite{Bourhis:1997yu}. In this paper,  the authors have proposed two fragmentation-function sets  BFGI and BFGII. The two sets differ primarily in the number of free parameters and  their values that are used in the parametrisation of the fragmentation functions when performing the fit to the data from ALEPH \cite{ALEPH:1995njw} and HRS \cite{Abachi:1989em}.  These differences mainly affect the gluon-to-photon fragmentation functions. In our study we will make use of both ALEPH and BFGII parametrisations of the photon-fragmentation functions.

%-----------------------------------------------------------
%
\section{Computational framework and modifications made}
\label{computational-setup}
%
%-----------------------------------------------------------

In our study we calculate tree-level and one-loop matrix elements  for the $pp \to e^+\nu_e \,\mu^- \nu_{\mu} \, b\bar{b} \, \gamma+X$ process with the help of the \textsc{Recola} matrix element generator \cite{Actis:2016mpe,Actis:2012qn} that we have modified to incorporate the random polarisation method \cite{Draggiotis:1998gr,Draggiotis:2002hm,Bevilacqua:2013iha}. This modification leads to significant improvements in the performance of the phase-space integration that is carried out with the help of \textsc{Parni} \cite{vanHameren:2007pt} and \textsc{Kaleu} \cite{vanHameren:2010gg}. Scalar and tensor one-loop integrals are calculated with \textsc{Collier} \cite{Denner:2016kdg}. In addition, we have implemented an alternative way of reducing 1-loop amplitudes to scalar integrals using \textsc{CutTools} \cite{Ossola:2007ax} and the OPP reduction method \cite{Ossola:2006us}. The resulting scalar integrals are further computed with \textsc{OneLOop} \cite{vanHameren:2010cp}.  We perform the OPP reduction and the evaluation of the scalar integrals  with quadruple precision, however, we calculate the tensor coefficients with double precision. Not only is this second reduction scheme used for additional cross-checks of our computational framework, but also for phase-space points containing tensor integrals that have been flagged as possibly unstable by \textsc{Collier}.  The singularities from soft and collinear parton emissions are isolated via subtraction methods for NLO QCD calculations. Specifically, we use the Nagy-Soper subtraction scheme \cite{Bevilacqua:2013iha}. However, we also cross-check our predictions against the results obtained with the commonly used Catani-Seymour dipole subtraction scheme \cite{Catani:1996vz,Catani:2002hc,Czakon:2009ss}. Both subtraction schemes are implemented in \textsc{Helac-Dipoles} \cite{Czakon:2009ss} that is part of the \textsc{Helac-Nlo} framework \cite{Bevilacqua:2011xh}. All additional technical details can be found in our previous papers dedicated to $pp\to t\bar{t}\gamma\,(\gamma) +X$  \cite{Bevilacqua:2018woc,Bevilacqua:2018dny,Stremmer:2023kcd,Stremmer:2024ecl}. 

In what follows, we will focus on the changes and modifications made in the  \textsc{Helac-Dipoles} Monte Carlo program, which are necessary to implement the photon fragmentation function in the real emission part of the full NLO QCD calculation. However, we would like to point out that a more comprehensive description can be found in Ref. \cite{Stremmer:992450}. While the QCD subtraction is basically identical for the different photon-isolation conditions and with respect to previous calculations of NLO QCD corrections to the $pp\to t\bar{t}\gamma$ process, the fixed-cone isolation condition allows additional collinear quark and photon configurations. Such configurations are handled with the modified $q\to q\gamma$ dipoles, which have to be differential with respect to the photonic momentum fraction, $\tilde{z}_{\gamma}$. The $\tilde{z}_{\gamma}$ variable from the dipole terms, that can be understood as a proxy for $z$ from the fragmentation function,  has to fulfil the following two conditions: in the collinear limit it should approach $\tilde{z}_{\gamma}\to z$, while for soft photons it should vanish. By using $\tilde{p}_{\gamma}=\tilde{z}_{\gamma}\,  \tilde{p}_{q\gamma}$ and $E_T^{\rm had}=(1-\tilde{z}_{\gamma})\, \tilde{p}_{T, \,q\gamma}$, the condition of the fixed-cone isolation approach given in Eq. \eqref{fixed-cone} is translated to a lower limit provided by $\tilde{z}_{\gamma}>z_{\rm cut} $,  where $z_{\rm cut}$ is a function depending on the transverse energy of $\tilde{p}_{\,q\gamma}$, whereas  $\tilde{p}_{q\gamma}$ is the auxiliary momentum of the  parent parton after the momentum mapping of the dipole subtraction scheme. Moreover, this condition must also be satisfied by $z$. In the calculation of the real subtracted part, the inclusion of the fixed-cone isolation condition can now be performed in a straightforward way. On the other hand, the calculation of the integrated dipoles have to be modified to retain the dependence on $\tilde{z}_{\gamma}$.

In the case of the Cataini-Seymour subtraction scheme the corresponding integrated dipoles have been calculated with an additional restriction on the phase space of the unresolved parton, parametrised  by the so-called $\alpha_{max}$ parameter. The corresponding results are provided in Ref. \cite{Denner:2014ina} for both massless final- and initial-state spectators. We have only implemented the case of final-state spectators and choose exactly one spectator particle $k$ to avoid the complications in the numerical evaluation coming from the additional plus distributions with respect to the momentum fractions of the initial-state partons for initial-state spectators. In this case the variable $\tilde{z}_{\gamma}$ is defined as 
\begin{equation}
\tilde{z}_{\gamma}=\frac{p_{\gamma}\cdot p_k}{p_q\cdot p_k+p_{\gamma}\cdot p_k}\,,
\end{equation}
and is directly connected to the integration variable $\tilde{z}_i=1-\tilde{z}_{\gamma}$ from  the original formulation of the Cataini-Seymour subtraction scheme, which simplifies the whole calculation. The integrated dipole is then given by
\begin{equation}
\begin{aligned}
\mathcal{V}_{q\gamma}^{\rm coll}(\alpha_{max},\epsilon,z_{\rm cut})&=Q_q^2\int_0^{1-z_{\rm cut}}d\tilde{z}_i\, (\tilde{z}_i(1-\tilde{z}_i))^{-\epsilon}\int_0^{\alpha_{max}}d y\, y^{-1-\epsilon} (1-y)^{1-2\epsilon}\\[0.2cm]
&\qquad\times\left[\frac{2}{1-\tilde{z}_i+y\tilde{z}_i}-(1+\tilde{z}_i)-\epsilon (1-\tilde{z}_i)\right]\\[0.2cm]
&=Q_q^2\int_0^{1-z_{\rm cut}}d\tilde{z}_i\,\bigg[\frac{1+\tilde{z}_i^2}{1-\tilde{z}_i}\left(-\frac{1}{\epsilon}+\log(\tilde{z}_i(1-\tilde{z}_i))+\log(\alpha_{max})\right)\\[0.2cm]
&+\alpha_{max}(1+\tilde{z}_i)+1-\tilde{z}_i-\frac{2}{\tilde{z}_i(1-\tilde{z}_i)}\log\left(\frac{1-\tilde{z}_i+\alpha_{max}\tilde{z}_i}{1-\tilde{z}_i} \right)  \bigg]\,,
\end{aligned}
\end{equation}
where the $\epsilon$ pole from the integrated dipoles cancels exactly the one in Eq. \eqref{eq:frag_ct}.

We write the local counterterms  in the Nagy-Soper subtraction scheme for general QED singularities as follows 
\begin{equation}
\begin{split}
\mathcal{A}^{D}& = \sum_{i,j,k=1}^{n+1}  
\mathcal{A}^{B}(\{\tilde{p}\}_n^{ij}) \otimes \mathcal{D}^{(ijk)}(\{\tilde{p}\}_n^{(ij)},\{p\}_{n+1})\\[0.2cm]
& =\sum_{i,j,k=1}^{n+1}  \,
\sum_{\tilde{s}_1,\tilde{s}_2= \pm}
\mathcal{A}_{\tilde{s}_1\tilde{s}_2}^{B}(\{\tilde{p}\}_n^{ij}) \, \mathcal{D}^{(ijk)}_{\tilde{s}_1\tilde{s}_2}(\{\tilde{p}\}_n^{(ij)},\{p\}_{n+1}) \, \left(\mathbf{Q}_{ij} \cdot \mathbf{Q}_k\right),
\end{split}
\end{equation}
where $i$ and $j$ label the splitting particles from $\widetilde{ij}\to i+j$, while the particle $k$ is the spectator. The Born matrix element before the splitting is denoted as $\mathcal{A}^{B}$, $\mathcal{D}^{(ijk)}$ are the splitting functions in the Nagy-Soper formalism and  $\{p\}_{n+1} \to \{\tilde{p}\}_n^{(ij)}$ relates the momenta before and after the splitting. The symbol $\otimes$ denotes spin correlations, while $\tilde{s}_1,\tilde{s}_2$ are the spin indices of the splitting particle $\tilde{p}_i$ and $\mathbf{Q}_{ij}$, $\mathbf{Q}_k$ are the corresponding charges with a relative minus sign between initial- and final-state particles. The  function $\mathcal{D}^{(ijk)}_{\tilde{s}_1\tilde{s}_2}$ can be further 
decomposed into two kind of contributions, $W^{(ii,j)}$ and  $W^{(ik,j)}$, called respectively diagonal and interference terms, according to 
\begin{equation} \label{eq:nsstruct}
\mathcal{D}^{(ijk)}_{\tilde{s}_1\tilde{s}_2}=W^{(ii,j)}_{\tilde{s}_1\tilde{s}_2}\delta_{ik}+W^{(ik,j)}_{\tilde{s}_1\tilde{s}_2}(1-\delta_{ik})\, \delta_{\tilde{s}_1\tilde{s}_2}\,.
\end{equation}
%}
The diagonal terms $W^{(ii,j)}_{\tilde{s}_1\tilde{s}_2}$ comprise both soft and collinear singularities, while the interference terms $W^{(ik,j)}_{\tilde{s}_1\tilde{s}_2}$ contain only soft singularities. The latter, however, are absent due to the lower cut on $\tilde{z}_{\gamma}$. Therefore, we can simply set $W^{(ik,j)}=0$. The calculation of the integrated-subtraction terms then closely follows the semi-numerical approach employed in Ref. \cite{Bevilacqua:2013iha}. We write the integration over the phase space of the unresolved parton as
\begin{equation}
D_{ii}=\int d e\,d c\,\Theta(\tilde{z}_{\gamma}(e,c)-z_{\rm cut})\, \frac{\mathcal{N}_{ii}^{d=4-2\epsilon}(e,c)}{(1-c)^{1+\epsilon}}\,,
\end{equation}
%,
where the integration variables $e$ and $c$, that are defined in Ref. \cite{Bevilacqua:2013iha},  describe the soft ($e\to 0$) and collinear ($c\to 1)$ limits. In addition, the pole structure of the integrand is explicitly factored out and the Heaviside function implements the lower cut of $\tilde{z}_{\gamma}(e,c)$, which is a function of the two integration variables. We can rewrite the integrated-subtraction term in the following way
\begin{equation} \label{eq:ns_dip}
\begin{aligned}
D_{ii}&=\int d e\,d c \frac{1}{(1-c)}\left[\mathcal{N}_{ii}^{d=4}(e,c)\Theta(\tilde{z}_{\gamma}(e,c)-z_{\rm cut})-\mathcal{N}_{ii}^{d=4}(e,1)\Theta(\tilde{z}_{\gamma}(e,1)-z_{\rm cut})\right]\\
&+\int d e\,d c \frac{1}{(1-c)^{1+\epsilon}}\left[\mathcal{N}_{ii}^{d=4-2\epsilon}(e,1)\Theta(\tilde{z}_{\gamma}(e,1)-z_{\rm cut})\right]\,,
\end{aligned}
\end{equation}
where, in the first line, we have subtracted a suitable counterterm that allows a numerical integration in $d=4$ dimensions. This counterterm is then added back in the second line,  where the only dependence on $c$ is found in the factor $1/(1-c)^{1+\epsilon}$, which after integration over $c$  leads to $-2^{-\epsilon}/\epsilon$. The remaining integration over $e$ is then again performed numerically. In the next step, we define the variable $\tilde{z}_{\gamma}(e,c)$ according to 
\begin{equation}
\tilde{z}_{\gamma}(e,c)=\frac{p_{\gamma}\cdot Q}{p_q\cdot Q+p_{\gamma}\cdot Q}=\frac{E_{\gamma}^{\rm CMS}}{E_{\gamma}^{\rm CMS}+E_{q}^{\rm CMS}}\,,
\end{equation}
where $Q$ denotes the total momentum of the process. This definition coincides with $\tilde{z}_{\gamma}$ in the Catani-Seymour subtraction scheme when  the following substitution is performed $p_k\to Q$. In the collinear limit ($c\to 1$) the photonic momentum fraction $\tilde{z}_{\gamma}(e,1)$ depends only on $e$,  which would allow us to invert this relation leading to $e(\tilde{z}_{\gamma})$ and rewrite the integration over $e$ in the second line of Eq. \eqref{eq:ns_dip} as the integration over $\tilde{z}_{\gamma}$. Instead, we rewrite the integration over $z_\gamma$ of the fragmentation functions as an integral over $e$. This allows us to combine the factorisation kernels in Eq. \eqref{eq:frag_ct} directly at the integrand level with the integrated subtraction terms so that all poles in $\epsilon$ cancel analytically before the integration over $e$ is performed.

%-----------------------------------------------------------
%
\section{Process definition and input parameters for LHC Run III}
\label{input-parameteres}
%
%-----------------------------------------------------------

\begin{figure}[t!]
    \begin{center}
	\includegraphics[trim = 18 724 18 18,width=1.0\textwidth]{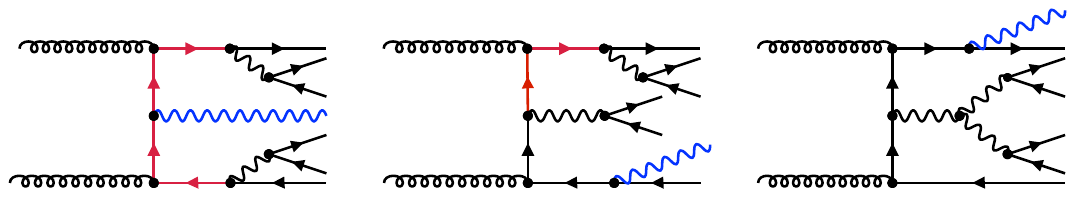}
    \end{center}
    \caption{\label{fig:FD} \it Representative Feynman diagrams, involving two (first diagram), one (second diagram) and no top-quark resonances (third diagram), contributing to  the $pp \to e^+\nu_e \,\mu^- \nu_{\mu} \, b\bar{b} \, \gamma+X$ process at leading order  defined at  ${\cal O}(\alpha_s^2\alpha^5)$. Red lines correspond to top quarks, blue lines to photons.} 
\end{figure}

We consider the $pp \to e^+\nu_e \,\mu^- \nu_{\mu} \, b\bar{b} \, \gamma+X$ process at LO and NLO in QCD  for  the LHC Run III center-of-mass energy of  $\sqrt{s}=13.6$ TeV.  Our calculations are based on the matrix elements for the final state $ e^+\nu_e \,\mu^- \nu_{\mu} \, b\bar{b} \, \gamma$  and include all resonant and  non-resonant Feynamn diagrams, their interferences and finite-width effects of the top quark and $W^\pm/Z$  boson. Thus, the propagators of unstable particles have the Breit-Wigner form. In addition, photons can be emitted from all the initial, final and intermediate (charged) particles. A few examples of Feynman diagrams contributing  to the $gg$ subprocess at ${\cal O}(\alpha_s^2\alpha^5)$ are presented in Figure \ref{fig:FD}. The calculation is performed in the so-called $5$-flavour scheme, which treats bottom quarks as massless  particles. In addition, the $5$-flavour scheme  resums initial state (potentially large) logarithms  in the bottom PDFs, which leads to more stable higher-order theoretical predictions.  For the LO contribution, the  $e^+\nu_e \,\mu^- \nu_{\mu} \, b\bar{b} \, \gamma$ final state can be produced via 
\begin{equation}
\begin{split}
gg & \to e^+\nu_e \,\mu^- \nu_{\mu} \, b\bar{b} \, \gamma \,,\\[0.2cm]
q\bar{q} & \to e^+\nu_e \,\mu^- \nu_{\mu} \, b\bar{b} \, \gamma \,,\\[0.2cm]
\bar{q}q & \to e^+\nu_e \,\mu^- \nu_{\mu} \, b\bar{b} \, \gamma \,,
\end{split}
\end{equation}
where $q$ stands for $q=u,d,c,s, b$. Since we require to have (at least) two $b$-jets of arbitrary charges in the final state, we must also include the following two subprocesses
\begin{equation}
\begin{split}
bb & \to e^+\nu_e \,\mu^- \nu_{\mu} \, bb \, \gamma \,,\\[0.2cm]
\bar{b}\bar{b} & \to e^+\nu_e \,\mu^- \nu_{\mu} \, \bar{b}\bar{b} \, \gamma \,.
\end{split}
\end{equation}
If we were able to precisely measure the charge of the two $b$-jets, then the last two subprocesses would not be needed. We note, however, that  the $b\bar{b}/\bar{b}b$ and $bb/\bar{b}\bar{b}$ initial-state contributions to the full  $pp \to e^+\nu_e \,\mu^- \nu_{\mu} \, b\bar{b} \, \gamma+X$  process are negligibly small and  at LO amount to only $0.2\%$ and $0.02\%$, respectively. We keep the Cabibbo-Kobayashi-Maskawa mixing matrix diagonal throughout the calculations. We use the NNPDF3.1 NLO PDF set \cite{NNPDF:2017mvq}  for both LO and NLO computations. The two-loop running of $\alpha_s$ is preformed with the help of the LHAPDF interface \cite{Buckley:2014ana}. The presence of the isolated photon in the final state requires a mixed scheme for the electromagnetic coupling constant $\alpha$, see e.g. Refs.  \cite{Bevilacqua:2018woc,Stremmer:2024ecl}. The total power of $\alpha$ is split into $\alpha^n=\alpha_{G_{\mu}}^{n-n_{\gamma}}\alpha (0)^{n_{\gamma}}$ where in our case $n_{\gamma}=1$. In particular, we use the $\alpha(0)$ scheme for the electromagnetic coupling associated with final-state photon radiation with $\alpha^{-1}(0) =137.035999084$ \cite{ParticleDataGroup:2022pth}, while for all other powers of $\alpha$ we use the $G_\mu$-scheme, where $\alpha_{G_\mu}$ is given by
\begin{equation}
	\alpha_{G_\mu} =\frac{\sqrt{2}}{\pi} \,G_\mu \, m_W^2\,\left(1-\frac{m_W^2}{m_Z^2}\right)\,,
	\quad \quad \quad \quad \quad \quad 
	G_{ \mu}=1.1663787 \cdot 10^{-5}~{\rm GeV}^{-2}\,.
\end{equation}
For the on-shell masses and widths of the $W^\pm/Z$ weak bosons we use the values from Ref.  \cite{ParticleDataGroup:2022pth}
\begin{equation}
\begin{array}{lll}
 m^{\rm OS}_{W}= 80.377 ~{\rm GeV}\,,&\quad\quad\quad\quad & \Gamma^{\rm OS}_{W} = 2.085 ~{\rm GeV}\,, 
\vspace{0.2cm}\\
 m^{\rm OS}_{Z}= 91.1876 ~{\rm GeV}\,,&\quad\quad\quad\quad & \Gamma^{\rm OS}_{Z} = 2.4955 ~{\rm GeV}\,,
\end{array}
\end{equation}
that are translated into their pole values according to the formulas given in Ref.  \cite{Bardin:1988xt} 
\begin{equation}
 m_V=\frac{m^{\textrm{OS}}_V}{\sqrt{1+\left(\Gamma^{\textrm{OS}}_V/m^{\textrm{OS}}_V\right)^2}} \,, \quad\quad\quad\quad \Gamma_V=\frac{\Gamma^{\textrm{OS}}_V}{\sqrt{1+\left(\Gamma^{\textrm{OS}}_V/m^{\textrm{OS}}_V\right)^2}}\,.
\end{equation}
  The full off-shell approach requires the evaluation of the top-quark width for unstable $W$ bosons. This calculation is based on the results presented in Refs. \cite{Denner:2012yc,Jezabek:1988iv}. Specifically, we use $\alpha_s(\mu_R=m_t)$ to compute NLO QCD corrections to $\Gamma_t$.  The corresponding LO and NLO top-quark widths  are given by
\begin{equation}
\begin{array}{lll}
 \Gamma_{t}^{\rm LO}= 1.4580658 ~{\rm GeV}\,,
 &\quad\quad\quad\quad & \Gamma_{t}^{\rm NLO}= 1.3329042~{\rm GeV}\,.
\end{array}
\end{equation}
The mass of the top quark is set to $m_t=172.5$ GeV, while all other fermions are considered massless. 

In the calculation with the fixed-cone isolation we use the ALEPH LO quark-to-photon fragmentation function. However, for comparison we also report results with the BFGII parton-to-photon  fragmentation functions. The BFGII set includes flavour-dependent quark-to-photon  and gluon-to-photon fragmentation functions. 
%Although the latter functions are not required in our case they are still included. 
The parametrisation of the ALEPH LO quark-to-photon fragmentation function is given in Eq. \eqref{eq:frag_aleph_lo}. The BFGII parton-to-photon fragmentation functions  are obtained from \textsc{Jetphox} \cite{Catani:2002ny}. To validate the implementation of the BFGII set in our system, we reproduced the results for the fragmentation contribution to the  $pp\to \gamma +1j$ process, which are presented in Ref. \cite{Chen:2022gpk}.  We note that, in the case of the ALEPH LO quark-to-photon fragmentation function, the dependence on the fragmentation scale is the same (up to a relative minus sign) for the direct and fragmentation contributions as defined in Eq. \eqref{eq:cross_pho_dir} and Eq. \eqref{eq:cross_pho_frag}, respectively. Thus, in this  case the full NLO result is independent of the fragmentation scale setting. The direct contribution  consists of the photon radiation at the matrix element level and the counterterm from the factorisation of the quark-to-photon fragmentation functions. The full dependence of the different parametrisations of the fragmentation functions is encoded in the fragmentation contribution, which is the convolution of the partonic $pp\to e^+\nu_e\,\mu^-\bar{\nu}_{\mu}\,b\bar{b}\,p'$ process with the parton-to-photon fragmentation functions, where $p'$ is either a gluon or a massless quark.

Based on the latest measurements of the $pp\to t\bar{t}\gamma$ process in the $\ell+jet$ and di-lepton decay channels performed by the ATLAS collaboration \cite{ATLAS:2024hmk} we employ a nested fixed-cone isolation, that comprises two fixed-cone photon isolation criteria. They are realised in our calculation by requiring that the transverse hadronic energy $E_{T}^{\rm had}$ inside the cone around the photon candidate with the radius $R=0.4$ is limited by
\begin{equation}
\label{isolation1}
    E_{T}^{\rm had} \le 0.022 \cdot p_{T, \, \gamma} + 2.45~{\rm GeV}\,,
\end{equation}
and within the smaller cone with radius $R=0.2$ by
\begin{equation}
\label{isolation2}
    E_{T}^{\rm had} \le 0.05 \cdot p_{T, \,\gamma}.
\end{equation}
The fixed-cone isolation with the smaller cone size has only a minor effect on our NLO QCD theoretical predictions. Indeed, when the results with just the outer fixed cone and the nested fixed cone isolation are compared  the integrated  cross section is reduced by less than $0.5\%$.  The calculations with the fixed-cone isolation are going to be compared with the predictions obtained using the smooth-cone and hybrid-photon isolation criteria. In these latter two cases the input parameters $(R,\varepsilon_{\gamma},n)$ of the (inner) smooth-cone isolation will be varied. In particular, for the smooth-cone isolation prescription we employ the following values  $\varepsilon_{\gamma}\in \{0.05,0.10,0.15,0.20,1.00\}$, $n\in \{0.5,1.0,2.0\}$ and $R=0.4$, whereas for the hybrid-photon isolation we will use instead $\varepsilon_{\gamma}\in \{0.05,0.10,0.15,0.20,1.00\}$, $n\in \{0.5,1.0,2.0\}$ and $R=0.1$.

We require the presence of at least two $b$-jets, two oppositely charged leptons and one photon. We closely follow the experimental environment in which it is difficult to determine the charge of the $b$-jet. Therefore,  in the recombination  of the partons in the jet algorithm the charges of bottom quarks are neglected. In practice, this means that the following recombination rules apply
\begin{equation}
bg\to b,\qquad \bar{b}g\to \bar{b},\qquad b\bar{b}\to g,\qquad bb\to g,\qquad \bar{b}\bar{b}\to g\,.
\end{equation}
This corresponds to the charge-blind $b$-jet tagging scheme as discussed in Ref. \cite{Bevilacqua:2021cit}. The $anti$-$k_T$ jet algorithm \cite{Cacciari:2008gp} with $R = 0.4$ is used to cluster partons into jets after the photon isolation criterion is applied. The fiducial phase-space of the prompt photon is defined according to
\begin{equation}
\begin{array}{lll}
 p_{T, \, \gamma}>20 ~{\rm GeV}\,,  
&\quad \quad\quad\quad\quad |y_{\gamma}|<2.37 \,.
&
\end{array}
\end{equation}
The $b$-jets and charged leptons have to pass the following requirements 
\begin{equation}
\begin{array}{lll}
p_{T,\,b}>25 ~{\rm GeV}\,,  
&\quad \quad\quad\quad\quad |y_b|<2.5 \,, 
 &\quad \quad\quad \quad \quad
\Delta R_{bb}>0.4\,, \\[0.2cm]
 p_{T,\,\ell}>25 ~{\rm GeV}\,,    
 &\quad \quad \quad \quad\quad|y_\ell|<2.5\,,&
\quad \quad \quad \quad \quad
\Delta R_{\ell
 \ell} > 0.4\,,\\[0.2cm]
\Delta R_{\ell\gamma}>0.4\,,  
&\quad \quad\quad\quad\quad \Delta R_{\ell b}>0.4 \,, 
&\quad \quad\quad\quad\quad
\Delta R_{b\gamma}>0.4\,.
\end{array}
\end{equation}
We set the renormalisation $(\mu_R)$ and factorisation $(\mu_F)$ scales to a common value $\mu_R=\mu_F=\mu_0$. Our default scale setting is given by
\begin{equation}
\mu_0=\frac{E_T}{4} \,, 
\end{equation}
where $E_T$ is defined according to
\begin{equation} \label{ttaa:scale_et}
E_T=\sqrt{m^2_{t}+p_{T, \,t}^2}+\sqrt{m^2_{t}+p_{T,\, \bar{t}}^2 }  + p_{T,\,\gamma}\,.
\end{equation} 
To build this scale setting the (anti)top-quark momentum has to be reconstructed. This can be achieved by minimising the following quantity
\begin{equation}
    \mathcal{Q}=|M(t)-m_t|+|M(\bar{t})-m_t|,
\end{equation}
where $M(t)$ and $M(\bar{t})$ are the invariant masses of the reconstructed top and anti-top quarks, respectively.  The latter are reconstructed from their decay products using the so-called resonance histories.  We assume that  the momenta of the leptons can be fully reconstructed, see e.g. Refs. \cite{Sonnenschein:2005ed,Raine:2023fko}. Since we do not tag the charge of the  $b$-jet, we can identify the following different resonance histories at LO 
\begin{equation}
\begin{array}{llll}
  (\mathrm{i}) \quad & t = W^+(\to e^+ \nu_e) \, b_1  &
\quad    \quad \quad \mathrm{and} \quad \quad \quad
  & \bar{t}  =W^-(\to \mu^- \bar{\nu}_\mu) \,  b_2 \,,\\[0.2cm]
  (\mathrm{ii}) \quad &   t = W^+( \to e^+ \nu_e ) \,b_1 \gamma &
 \quad \quad \quad \mathrm{and}  \quad \quad \quad
  & \bar{t} = W^-(\to \mu^- \bar{\nu}_\mu)\,  b_2 \,,\\[0.2cm]
  (\mathrm{iii}) \quad &  t = W^+( \to e^+ \nu_e) \,b_1 &
\quad       \quad  \quad\mathrm{and}\quad \quad\quad \quad\quad
  & \bar{t} = W^-(\to \mu^- \bar{\nu}_\mu)\,  b_2 \gamma\,,\\[0.2cm]
\end{array}
\end{equation}
plus additional three cases that can be obtained by simply replacing $b_1\leftrightarrow b_2$.  In general these $6$ categories are not sufficient if NLO QCD calculations are considered. In particular, we can encounter events with up to three $b$-jets in the final state. In this case the total number of $18$  resonance histories have to be considered. Finally, an additional light jet (if resolved by the $anti$-$k_T$ jet algorithm and passed all the cuts) is not considered in the reconstruction.

In addition to the default scale choice, we also provide selected results for the following two scale settings 
\begin{equation} \label{ttaa:scale_mt}
\mu_0=m_t \,,
\end{equation}
and 
\begin{equation} \label{ttaa:scale_ht}
\mu_0=\frac{H_T}{4}\,,
\end{equation}
where $H_T$ is given by  
\begin{equation}
H_T=p_{T, \, b_1}+p_{T, \, b_2}+p_{T, \, e^+}+p_{T,\, \mu^-}+p_{T}^{miss}+p_{T, \, \gamma}\,.
\end{equation}
The missing transverse momentum is defined as $p_T^{miss} = |\vec{p}_{T,\, \nu_e} + \vec{p}_{T,\, \bar{\nu}_\mu}|$. These two scale settings do not require the reconstruction of the (anti)top-quark momentum, which can become ambiguous in the full off-shell calculation due to single- and non-resonant contributions, where only one or even zero resonant top quarks are present.   Theoretical uncertainties due to missing higher-order corrections are estimated based on a $7$-point scale variation, where $\mu_R$ and $\mu_F$ are changed independently in  the range
\begin{equation}
\frac{1}{2} \, \mu_0  \le \mu_R\,,\mu_F \le  2 \,  \mu_0\,, \quad \quad 
\quad \quad \quad \quad \quad \quad \frac{1}{2}  \le
\frac{\mu_R}{\mu_F} \le  2 \,.
\end{equation}
This  results in the following pairs that need to be evaluated
\begin{equation}
\label{scan}
\left(\frac{\mu_R}{\mu_0}\,,\frac{\mu_F}{\mu_0}\right) = \Big\{
\left(2,1\right),\left(0.5,1  
\right),\left(1,2\right), (1,1), (1,0.5), (2,2),(0.5,0.5)
\Big\} \,.
\end{equation}
The scale uncertainties are obtained by selecting the maximum and minimum values from the obtained results. Any uncertainties due to the different parametrisations of the parton-to-photon fragmentation functions
or the choice of the fragmentation scale in the case of the BFGII set are not taken into account in the NLO QCD
calculation due to the small relative size of the fragmentation contribution. Finally, we do not study the PDF uncertainties for the $pp\to e^+\nu_e \, \mu^-\bar{\nu}_\mu \, b\bar{b} \, \gamma+X$ process as they have already been examined in depth in Refs. \cite{Bevilacqua:2018woc,Bevilacqua:2019quz}. 

%
%-----------------------------------------------------------
%
\section{Integrated cross-section results for various photon isolation criteria}
\label{comparison-integrated}
%
%-----------------------------------------------------------
We start presenting the results for the $pp \to e^+\nu_e\,\mu^-\bar{\nu}_{\mu}\,b\bar{b}\,\gamma +X$ process at the LHC with $\sqrt{s}=13.6$ TeV by assessing the magnitude of the fragmentation contribution. Our NLO QCD results, together with their theoretical uncertainties related to the scale variation, are summarised in Table \ref{tab:tta_frag_functions}. They are obtained with the fixed-cone isolation prescription, by employing the ALEPH LO and BFGII parton-to-photon fragmentation functions. Separately, we also show only the fragmentation contribution for both parametrisations. The last column displays the ratio of the results obtained with the ALPEH LO quark-to-photon fragmentation function to the BFGII parton-to-photon fragmentation functions. Following the findings of Ref. \cite{Catani:2013oma}, we set the fragmentation scale to
\begin{equation}
\mu_{Frag}=R\,p_{T,\,\gamma}\,, 
\end{equation}
where  $R=0.4$.
\begin{table*}[t!]    
    \centering
    \renewcommand{\arraystretch}{1.5}
    \begin{tabular}{l@{\hskip 10mm}l@{\hskip 10mm}l@{\hskip 10mm}c}
        \hline\noalign{\smallskip}
        & ALEPH LO & BFGII& ALEPH LO / BFGII \\
        \noalign{\smallskip}\midrule[0.5mm]\noalign{\smallskip}
        $\sigma^{\rm NLO}$ [fb] & $ 21.50(2)^{+1.4\%}_{-5.0\%} $ & $ 21.48(2)^{+1.4\%}_{-5.0\%} $ & $ 1.001 $ \\
        $\sigma^{\rm NLO}_{\rm frag}$ [fb] & $ 0.038724(8) $ & $ 0.017020(4) $ & $ 2.275 $ \\
        \noalign{\smallskip}\hline\noalign{\smallskip}
    \end{tabular}
    \caption{\label{tab:tta_frag_functions} \it Integrated cross sections at NLO in QCD for the $pp\to e^+\nu_e\,\mu^-\bar{\nu}_{\mu}\,b\bar{b}\,\gamma +X$ process at the LHC with $\sqrt{s}=13.6~{\rm TeV}$. Theoretical results are obtained with the fixed-cone isolation prescription. The ALEPH LO and the BFGII parton-to-photon fragmentation functions are employed. Results only with the fragmentation contribution are also shown. Predictions are provided for 
    $\mu_0=E_T/4$ with the NNPDF3.1 NLO PDF set. Also displayed are the theoretical uncertainties coming from a $7$-point scale variation ($\pm$ percentages) and Monte Carlo integration errors  (in brackets).}
\end{table*}
We note once again that the sum of the direct and the fragmentation contribution is independent of $\mu_{Frag}$ for the ALEPH LO quark-to-photon fragmentation function. In this case, the fragmentation scale is only relevant if the fragmentation contribution is shown separately. We can observe that the ALEPH LO fragmentation function yield a $130\%$  larger fragmentation contribution compared to the result obtained with the BFGII set. This large difference in part is due to the inner cone of $R=0.2$  in the fixed-cone isolation. Once  the inner cone is removed the difference is reduced to $80\%$.  Similar large differences between these two parametrisation have already been presented in the literature, for example for  the $pp \to \gamma + jet$ process \cite{Schuermann:2022qdm}. In this  work  (N)NLO integrated and differential cross sections are provided for $pp \to \gamma + jet$ with the fixed-cone isolation including both, direct and fragmentation contributions. These large differences originate mainly from the different scale evolution of the quark-to-photon fragmentation functions and can be reduced when the fixed-order NLO quark-to-photon fragmentation function, first determined in Ref. \cite{Gehrmann-DeRidder:1997fhc,Gehrmann-DeRidder:1997fom}, is used instead of the ALEPH LO set. 
\begin{figure}[t!]
    \begin{center}
	\includegraphics[width=0.75\textwidth]{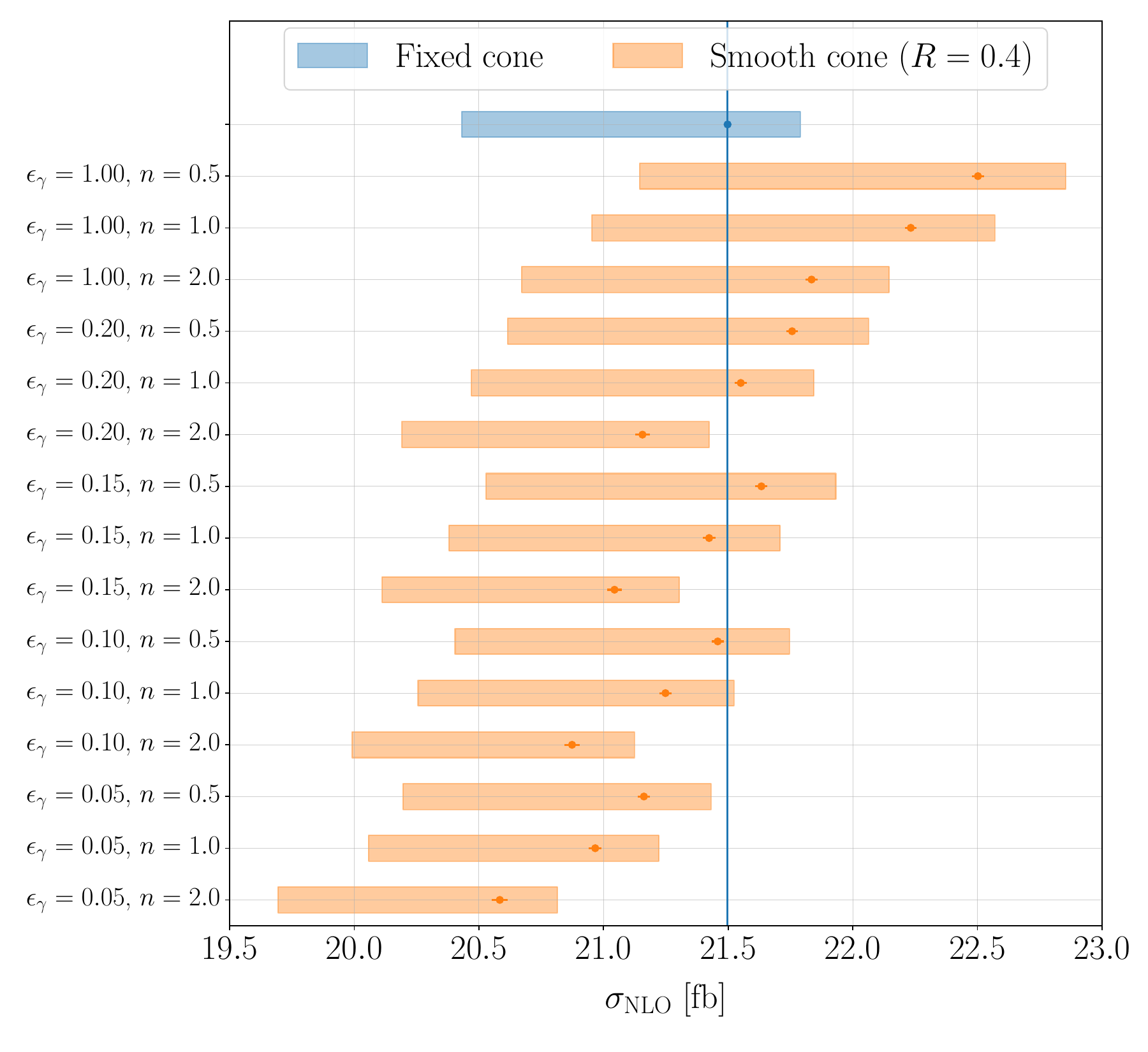}
    \end{center}
\caption{\label{fig:tta_frag_comp_smooth} \it Comparison of integrated  cross sections at NLO QCD for the $pp\to e^+\nu_e\,\mu^-\bar{\nu}_{\mu}\,b\bar{b}\,\gamma+X$ process at the LHC with $\sqrt{s}=13.6~{\rm TeV}$. Theoretical results are obtained using the fixed-cone and smooth-cone isolation prescription. For the results obtained with the smooth-cone isolation the following parameters are employed $\varepsilon_{\gamma}\in \{0.05,0.10,0.15,0.20,1.00\}$, $n\in \{0.5,1.0,2.0\}$ and $R=0.4$. Predictions are provided for $\mu_0=E_T/4$ with the NNPDF3.1 NLO PDF set.  Also displayed  (as bands) are the theoretical uncertainties coming from a $7$-point scale variation  and  (as bars) Monte Carlo integration errors.} 
\end{figure}
\begin{figure}[t!]
    \begin{center}
	\includegraphics[width=0.75\textwidth]{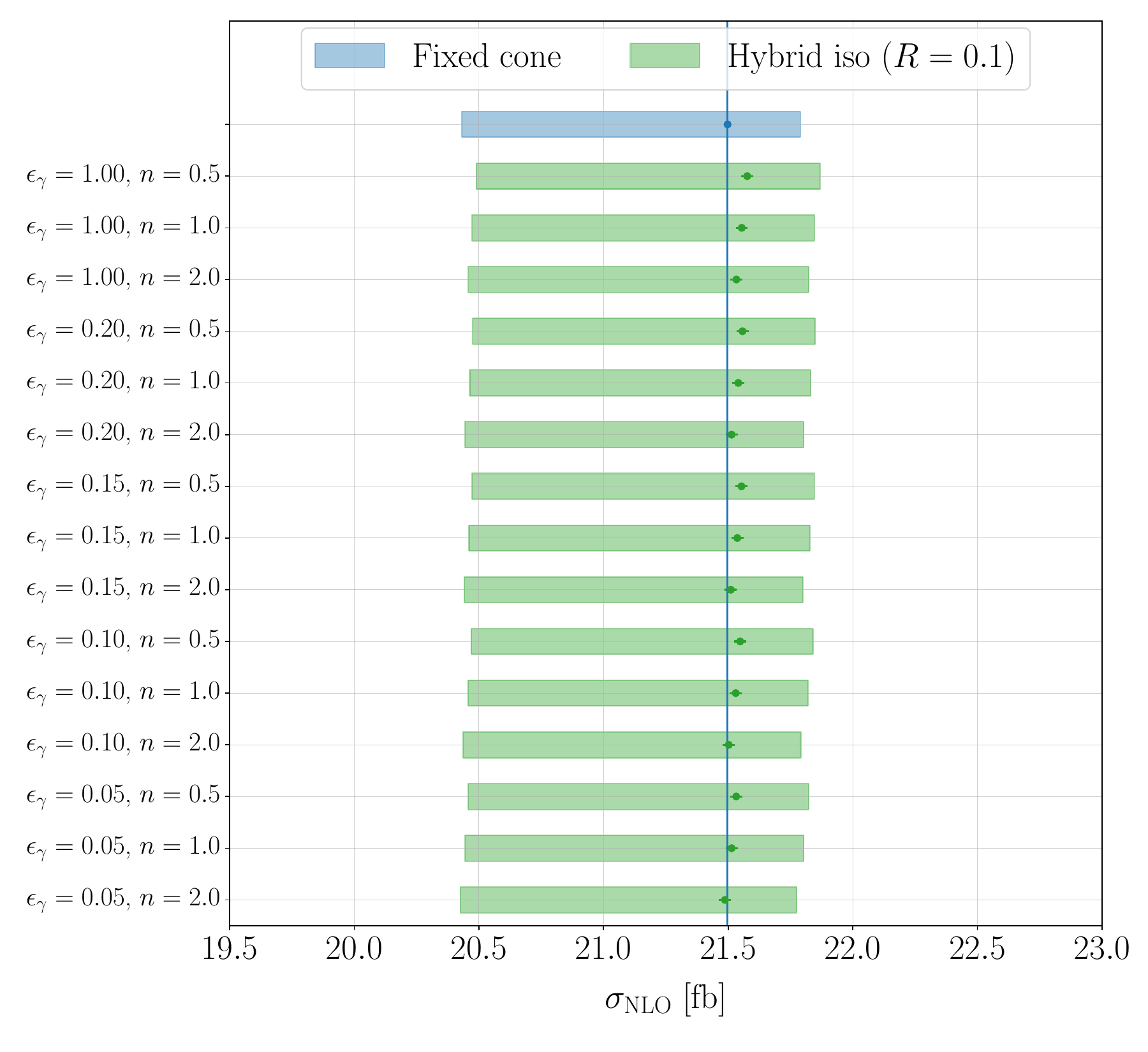}
    \end{center}
    \caption{\label{fig:tta_frag_comp_hybrid} \it Same as Figure \ref{fig:tta_frag_comp_smooth} but for the hybrid-photon isolation prescription with the following parameters $\varepsilon_{\gamma}\in \{0.05,0.10,0.15,0.20,1.00\}$, $n\in \{0.5,1.0,2.0\}$ and $R=0.1$ instead of the smooth-cone isolation prescription. }
\end{figure}

The contribution from the fragmentation process to the NLO integrated cross section for the $pp\to e^+\nu_e\,\mu^-\bar{\nu}_{\mu}\,b\bar{b}\,\gamma +X$ process is less than $0.2\%$ $(0.1\%)$ for the ALEPH LO quark-to-photon fragmentation function (for the BFGII set). Thus,  it is negligibly small compared to the theoretical uncertainties due to scale dependence, which  for this process are of the order of $5\%$. Generally, the fragmentation contribution highly depends on the actual event selection and the photon isolation criterion, but in this particular case its small size is directly related to the absence of light jets in the LO contribution. Since both $b$-jets in the final state have to be resolved, thus, can not fragment into a photon, the fragmentation contribution in the $gg$ and $q\bar{q}/\bar{q}q$ channels is strongly suppressed. In these two cases only an additional gluon can fragment into a photon. However, the corresponding gluon-to-photon fragmentation is included only in the BFGII set. As a consequence, the main contribution comes from the $qg/\bar{q}g$ subprocesses, where an additional light quark in the final state can fragment into a photon. In addition, in the PDF-suppressed $bg/\bar{b}g$ production modes, a third bottom quark, if present in the final state, can fragment into a photon. It should be noted that this situation will change in the $\ell+jet$ decay channel, where the fragmentation contribution will increase because of the presence of the two light quarks from the decay of the $W$ gauge boson.  For this decay channel, the $gg$ and $q\bar{q}/\bar{q}q$ production modes will no longer vanish for the ALEPH LO quark-photon fragmentation function.

In the next step, a comparison between the calculation with the fixed-cone isolation and alternative predictions obtained with the smooth-cone isolation is performed. In Figure \ref{fig:tta_frag_comp_smooth} we show the NLO QCD results with their corresponding scale uncertainties for the fixed-cone and smooth-cone isolation, where in the latter case we set $R=0.4$ and vary the other two free parameters, $(\varepsilon_{\gamma},\, n)$, in the following ranges $\varepsilon_{\gamma}\in \{0.05,0.10,0.15,0.20,1.00\}$ and $n\in \{0.5,1.0,2.0\}$. The largest differences between the two isolation criteria are found for $(\varepsilon_{\gamma},n)=(1.00,0.5)$ and $(\varepsilon_{\gamma},\, n)=(0.05,2.0)$. They are of the order of  $4.7\%$ and $4.3\%$, respectively. These differences are similar in size to the NLO QCD scale uncertainties, which are at the $5\%$ level, and are therefore not negligible. The result obtained with the smooth-cone isolation for $(\varepsilon_{\gamma},\,n)=(1.00,1.0)$, that is often used in higher-order calculations involving photons, differs from the prediction with the fixed-cone isolation by about $3.4\%$. This difference is still significant  compared to the size of the scale uncertainties. The normalisation differences seen in Figure \ref{fig:tta_frag_comp_smooth} can be avoided by tuning the input parameters. Indeed, they can be reduced to the negligible $0.2\%-0.4\%$ level if one chooses a single configuration from the following set  $(\varepsilon_{\gamma}, \,n)=\left\{(0.20,1.0), \, (0.15,1.0), \, (0.10,0.5)\right\}$, instead of the commonly used values. However, such tuning is not only very impractical, but also time-consuming. Firstly, the cross section for the  $pp \to e^+ \nu_e \, \mu^- \bar{\nu}_\mu \, b\bar{b}\,\gamma+X$ process is very sensitive to the specific input values used for $(\varepsilon_{\gamma}, \,n)$. Indeed, the differences of up to even $9\%$ can be observed between the maximal and minimal cross-section values. Secondly, the tuning procedure depends on the process and decay channel under consideration as well as the phase-space cuts used. Since the sensitivity to input parameters also increases with the presence of more photons and jets in the final state, the tuning procedure would have to be repeated in each case separately.

Instead of relying entirely on the smooth-cone isolation, it is also possible to use the hybrid-photon isolation.  In this case, in the first step, the smooth-cone isolation is used in a smaller cone $R$ located within the cone size $R_{\rm fixed}$, such that $R < R_{\rm fixed}$. After that, the outer cone of size $R_{\rm fixed}$ is used to apply the experimental fixed-cone isolation with the exact parameters as in the experimental analysis.  In addition, this approach has the advantage that the dependence on the input parameters $(\varepsilon_{\gamma}, \,n)$ of the inner smooth-cone isolation is reduced as presented in Figure \ref{fig:tta_frag_comp_hybrid}, where a comparison between the NLO QCD results with the fixed-cone and hybrid-photon isolation is given. Also displayed are the corresponding theoretical uncertainties coming from a $7$-point scale variation. As in the previous case the parameters $(\varepsilon_{\gamma}, \,n)$ are varied in the same ranges, but this time  
we set $R=0.1$.  We can observe that the dependence of the inner smooth-cone isolation on the input parameters is negligibly small. Indeed, the differences are only up to $0.4\%$ regardless of the values used for  $(\varepsilon_{\gamma}, \,n)$. Furthermore, all NLO QCD predictions obtained with the hybrid-photon isolation agree very well with the result calculated with the fixed-cone isolation. The largest differences we have noticed are only of the order of $0.4\%$, thus, well within the corresponding uncertainties. We have varied the size of the inner radius in the smooth-cone isolation method to examine the dependence of our results on $R$.  In particular, the spread of the predictions obtained with the hybrid-photon isolation with different input parameters changes from about $0.4\%$ for $R=0.1$ to $0.7\%$ for $R=0.15$ and to $1.0\%$ for $R=0.2$, while the other parameters are again varied in the same ranges as before. Accordingly, the hybrid-isolation can be safely used instead of the fixed-cone isolation for the $pp \to e^+ \nu_e \, \mu^- \bar{\nu}_\mu \, b\bar{b} \, \gamma +X$ process,  since the largest differences between the two photon isolation criteria are less than $0.4\%$ for $R=0.15$ and less than $0.7\%$ for $R=0.2$. We also observe that  the dependence on $R$ increases for decreasing $\varepsilon_{\gamma}$ and increasing $n$. We find that the dependence on $R$ is at the per mille level  for the three values $R\in \{0.10,0.15,0.20\}$ when $(\varepsilon_{\gamma},n)=(1.00,0.5)$. On the other hand, for $(\varepsilon_{\gamma},n)=(0.05,2.0)$ the calculations for the three values of $R$ differ by about $1\%$.  Finally, we note that for all the results presented in Figure \ref{fig:tta_frag_comp_hybrid} the scale uncertainties are identical. To summarise this part, the dependence on the input  parameters of the smooth-cone isolation is greatly reduced for the hybrid-photon isolation prescription. In each case, and without any tuning of these input parameters, good agreement with the calculations obtained for the fixed-cone insolation can be clearly observed.

%
%-----------------------------------------------------------
%
\section{Differential cross-section results for various photon isolation criteria}
\label{comparison-differential}
%
%-----------------------------------------------------------
%
\begin{figure}[t!]
    \begin{center}
	\includegraphics[width=0.49\textwidth]{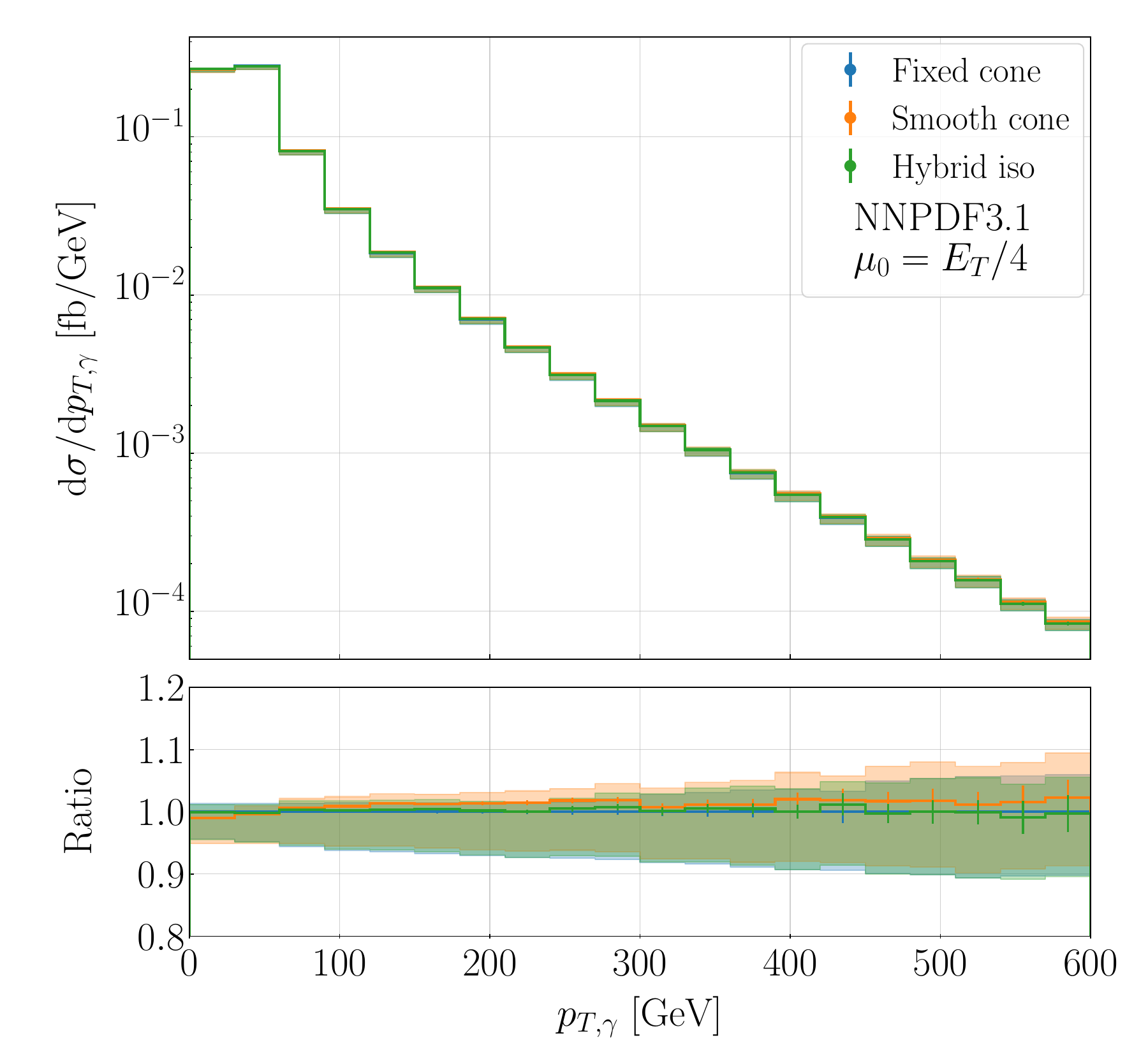}
	\includegraphics[width=0.49\textwidth]{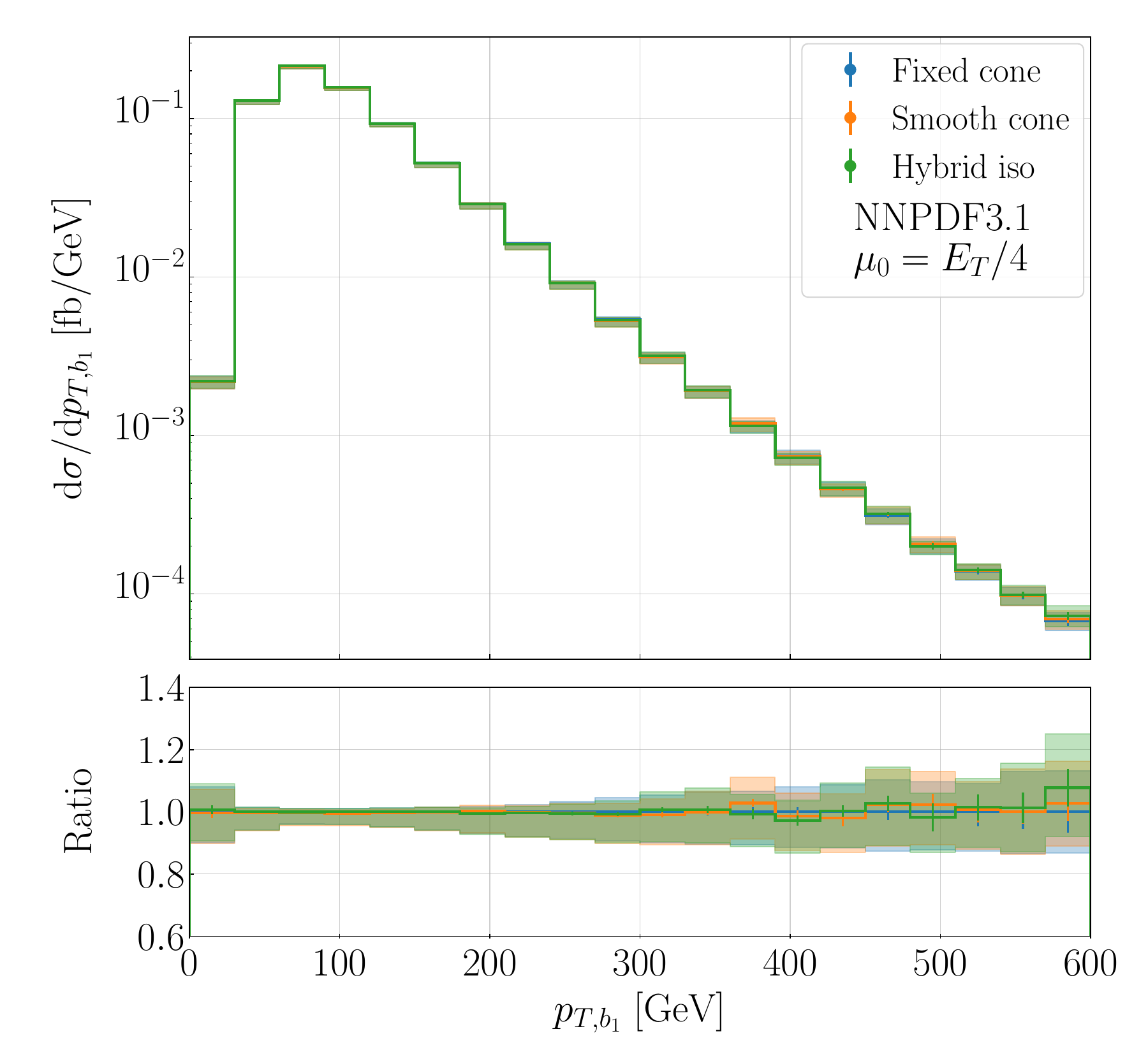}
	\includegraphics[width=0.49\textwidth]{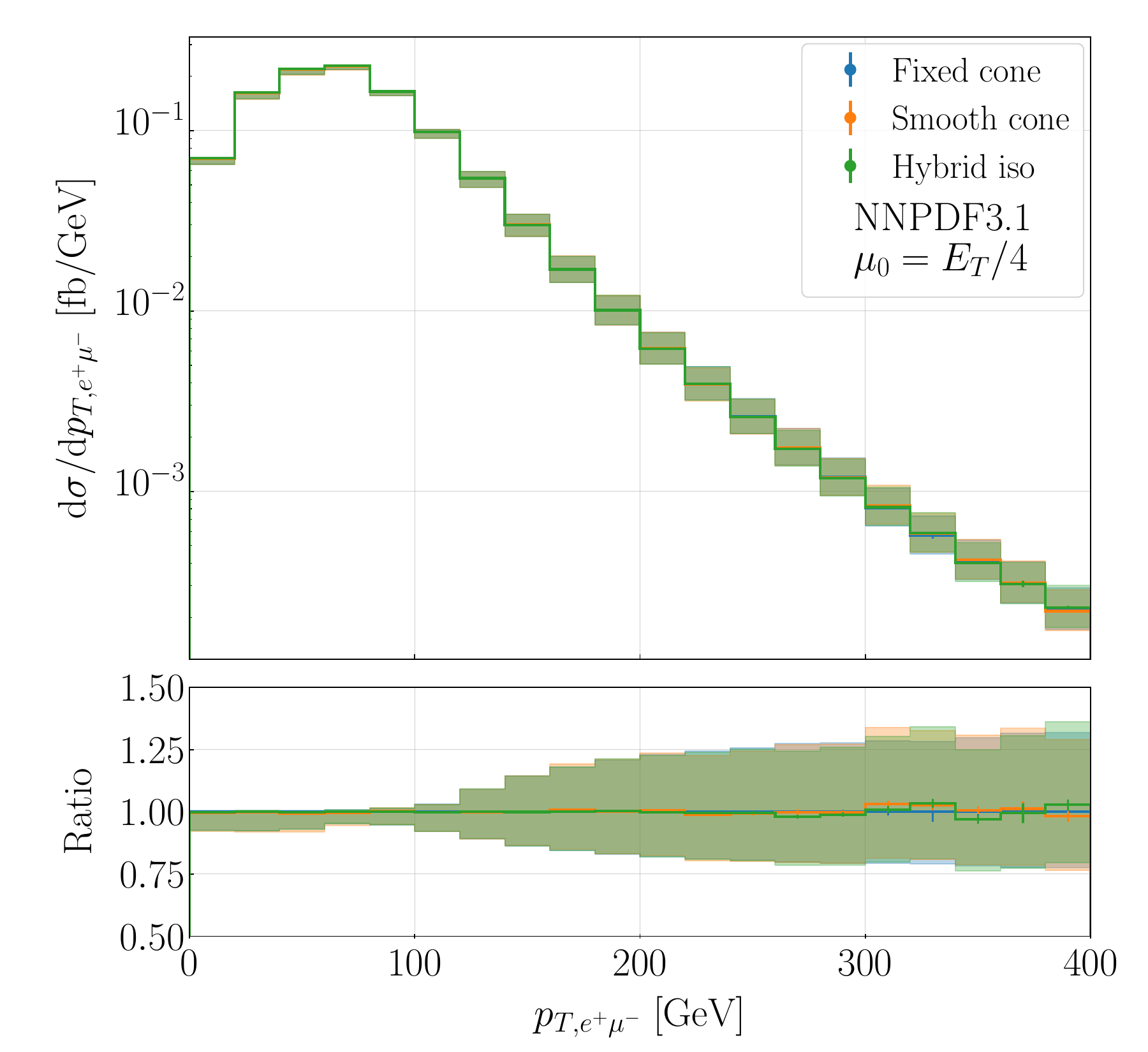}
    \end{center}
    \caption{\label{fig:tta_frag_phot_iso1} \it Differential  cross-section distributions at NLO in QCD  for the $pp\to e^+\nu_e\,\mu^-\bar{\nu}_{\mu}\,b\bar{b}\,\gamma+X$ process at the LHC with $\sqrt{s}=13.6$ TeV  as a function of  $p_{T,\,\gamma}$, $p_{T,\,b_1}$ and $p_{T,\,e^+\mu^-}$. Results are presented for the fixed-cone isolation, smooth-cone isolation with $(\epsilon_{\gamma},n)=(0.10,0.5)$ and hybrid-photon isolation with $(\epsilon_{\gamma},n)=(0.10,2.0)$. The lower panels show the ratio to the result obtained with the fixed-cone isolation. Also displayed (as bands) are the theoretical uncertainties coming from a $7$-point scale variation  and (as bars) Monte Carlo integration errors. The scale choice is set to  $\mu_0=E_T/4$. The cross sections are evaluated with the NNPDF3.1 NLO PDF set. }
\end{figure}
\begin{figure}[t!]
    \begin{center}
	\includegraphics[width=0.49\textwidth]{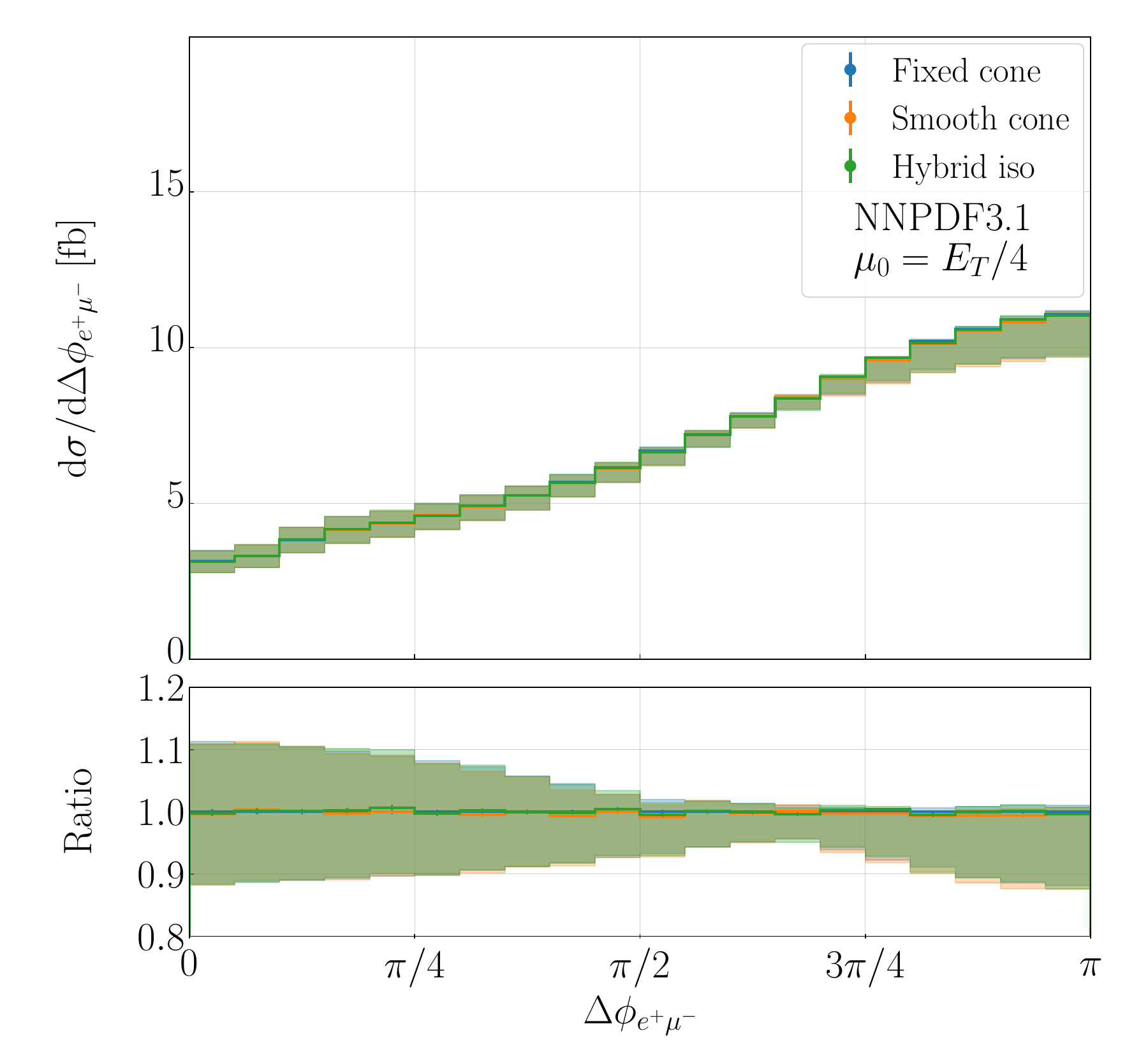}
 	\includegraphics[width=0.49\textwidth]{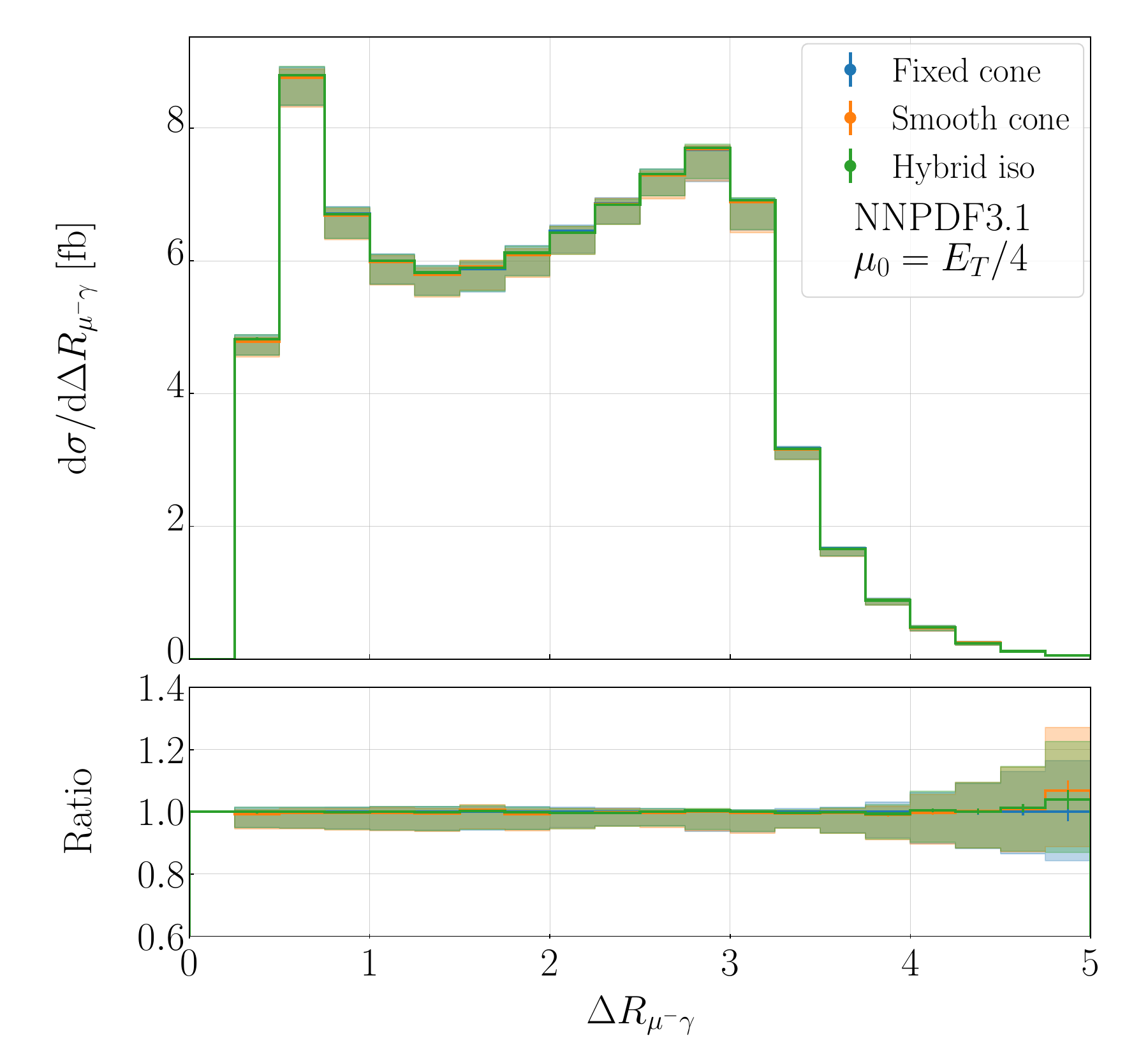}
	\includegraphics[width=0.49\textwidth]{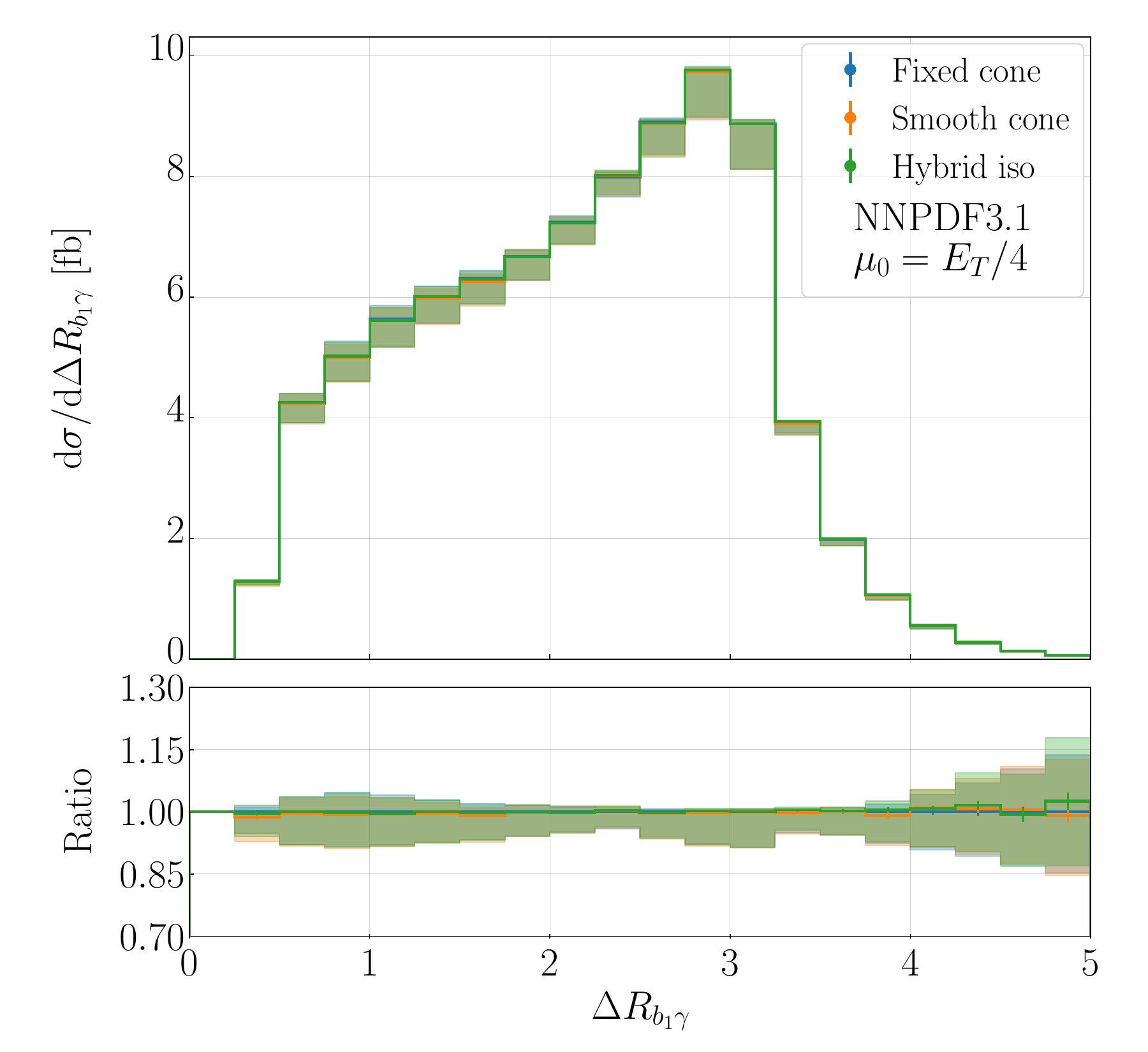}
    \end{center}
    \caption{\label{fig:tta_frag_phot_iso2} \it Same as Figure \ref{fig:tta_frag_phot_iso1} but for  $\Delta \phi_{e^+\mu^-}$, $\Delta R_{\mu^-\gamma}$ and $\Delta R_{b_1\gamma}$. }
\end{figure}

In this section, we examine the impact of applying three photon-isolation criteria on various differential cross-section distributions. Based on our findings from the previous section, we choose the following parameters for the smooth-cone isolation approach $(\varepsilon_{\gamma},n)=(0.10,0.5)$ and $R=0.4$. At the integrated cross-section level, this set of the parameters  has led to differences  of only $0.3\%$ compared to the prediction obtained with the fixed-cone isolation. The parameters of the hybrid-photon isolation and in particular of the inner smooth-cone isolation are set to $(\epsilon_{\gamma},n)=(0.10,2.0)$ with $R=0.1$. Overall, we can say that for all the observables we have studied, the predictions for the three different photon-isolation criteria agree very well. As an example, in Figure \ref{fig:tta_frag_phot_iso1} we present the results for the following dimensionful observables: the transverse momentum of the photon $(p_{T,\, \gamma})$, the hardest $b$-jet $(p_{T, \, b_1})$ and the pair of two charged leptons $(p_{T,\, e^+\mu^-})$. On the other hand, in Figure \ref{fig:tta_frag_phot_iso2} we display the following dimensionless observables: the azimuthal-angle difference between the two charged leptons $(\Delta \phi_{e^+\mu^-})$, the angular separation in the $\phi-y$ plane between the muon and the photon $(\Delta R_{\mu^-\gamma})$ as well as the hardest $b$-jet and the photon $(\Delta R_{b_1\gamma})$. In each case the upper panels present the absolute NLO QCD predictions together with their corresponding theoretical uncertainties from a $7$-point scale variation estimated on a bin-by-bin basis. The lower panels show the ratio to the result obtained with the fixed-cone isolation prescription. All the results presented in this section are calculated for $\mu_R=\mu_F=\mu_0=E_T/4$ and with the NNPDF3.1 NLO PDF set. 

For all three results obtained using the fixed-cone, smooth-cone (with the tuning) and hybrid-photon isolation, we can observe that the differences are very small indeed compared to the respective scale uncertainties. Only for $p_{T,\,\gamma}$ we find that the prediction with the smooth-cone isolation is consistently larger than the predictions obtained with the other two photon isolation criteria. This can be observed over the entire plotted range. However, these differences amount to only $1\%-2\%$, so they are negligible compared to the corresponding scale uncertainties, which are at the level of $5\%-10\%$.
\begin{figure}[t!]
    \begin{center}
    \includegraphics[width=0.49\textwidth]{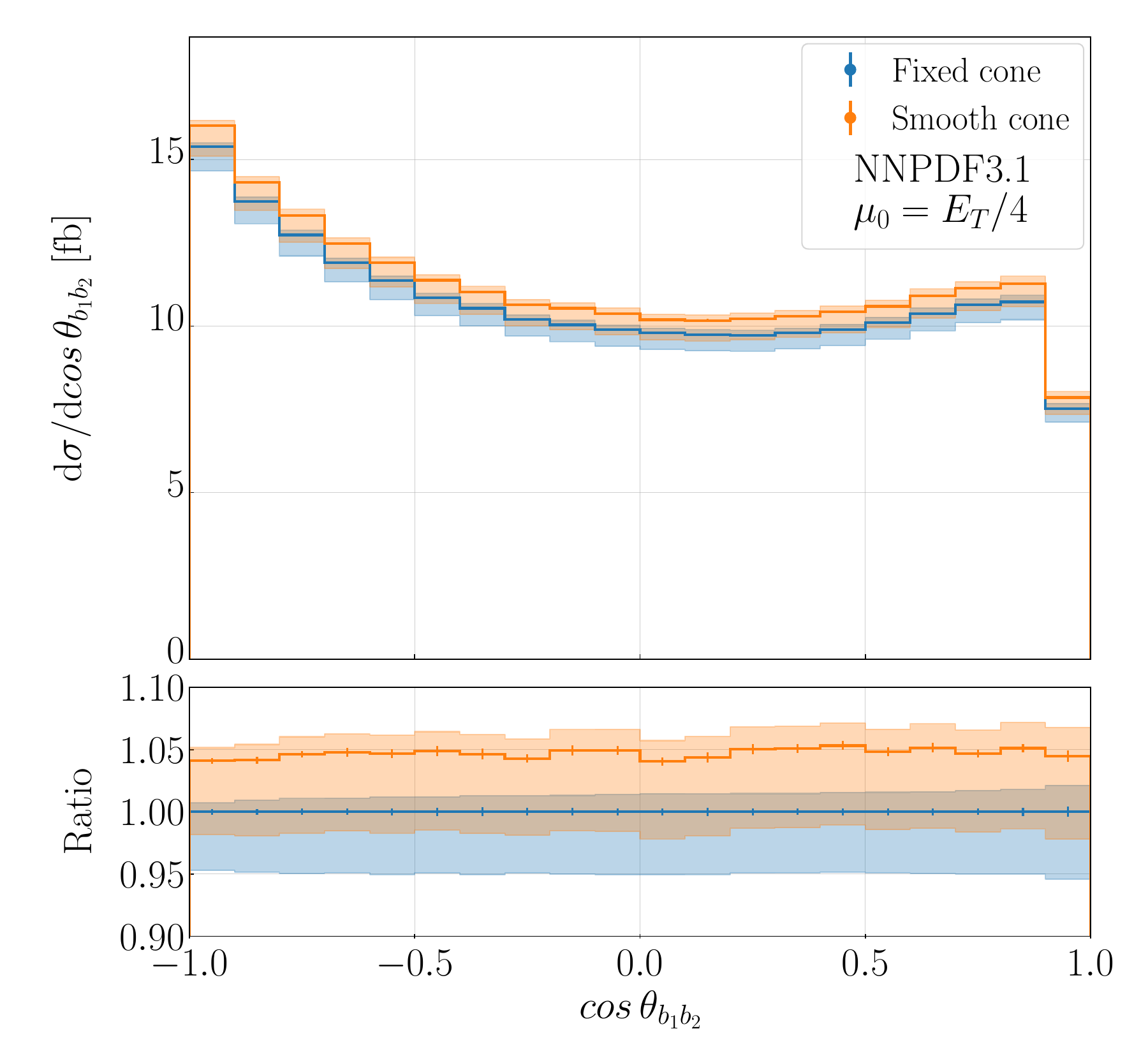}
	\includegraphics[width=0.49\textwidth]{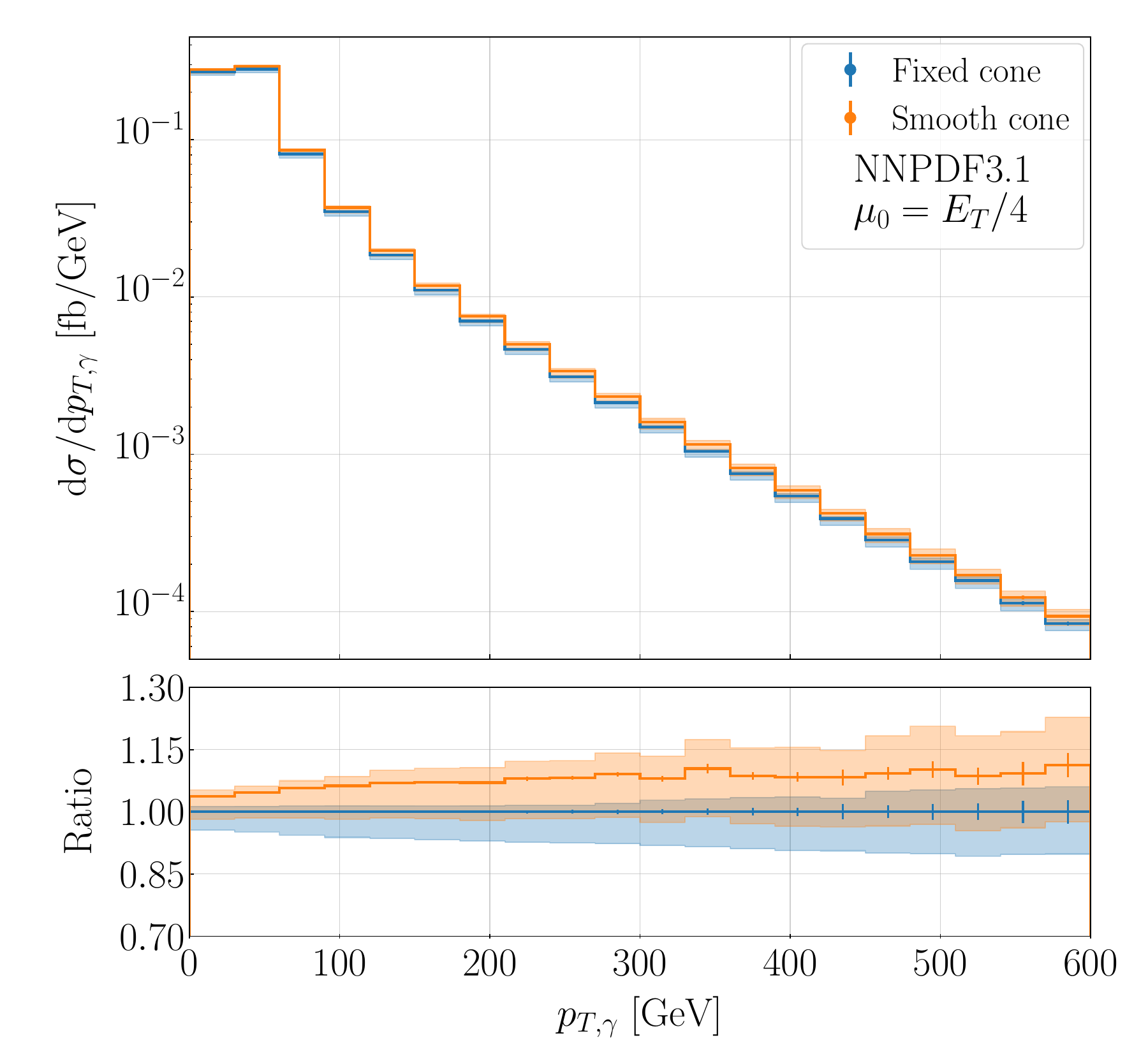}
	\includegraphics[width=0.49\textwidth]{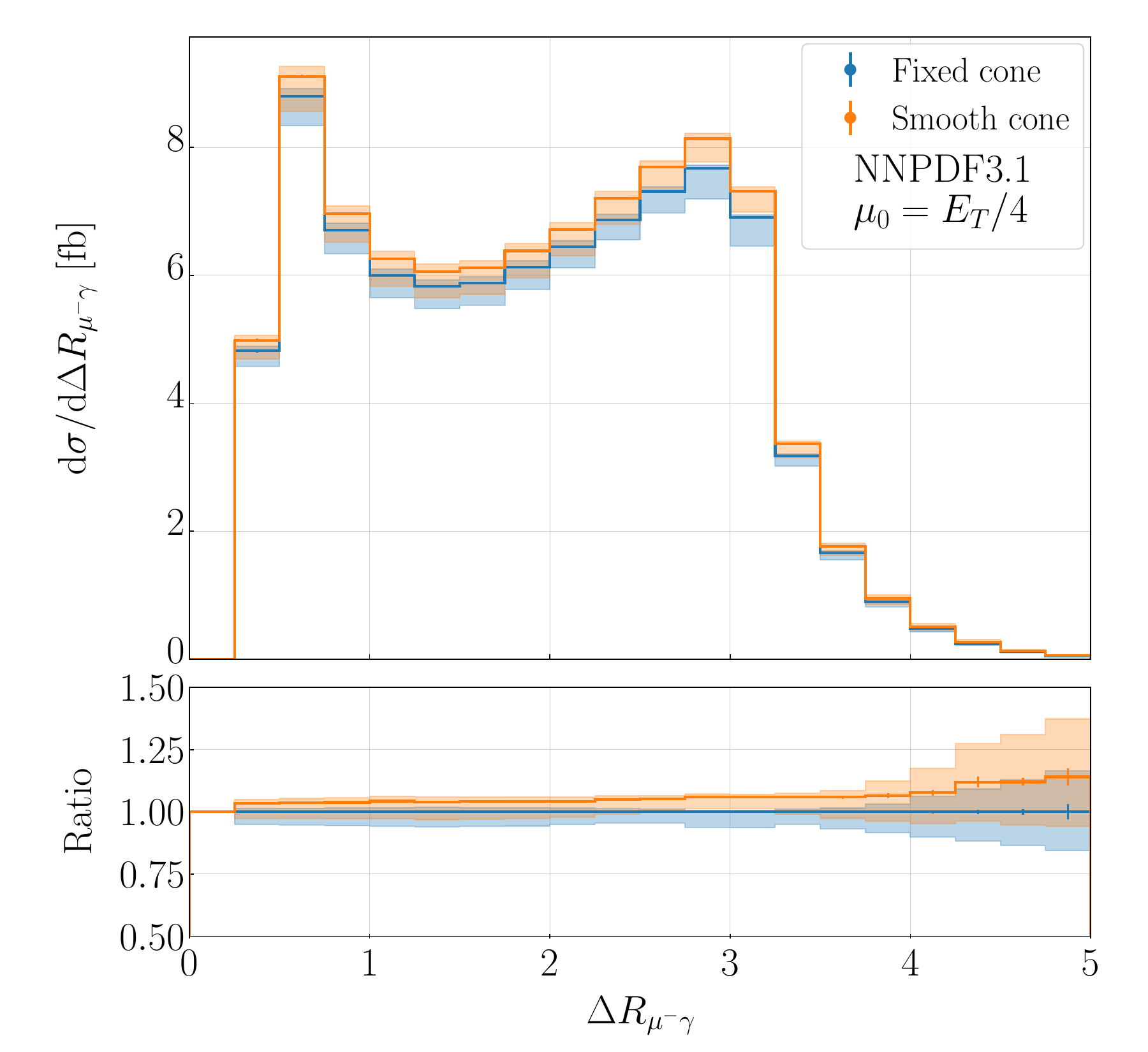}
	\includegraphics[width=0.49\textwidth]{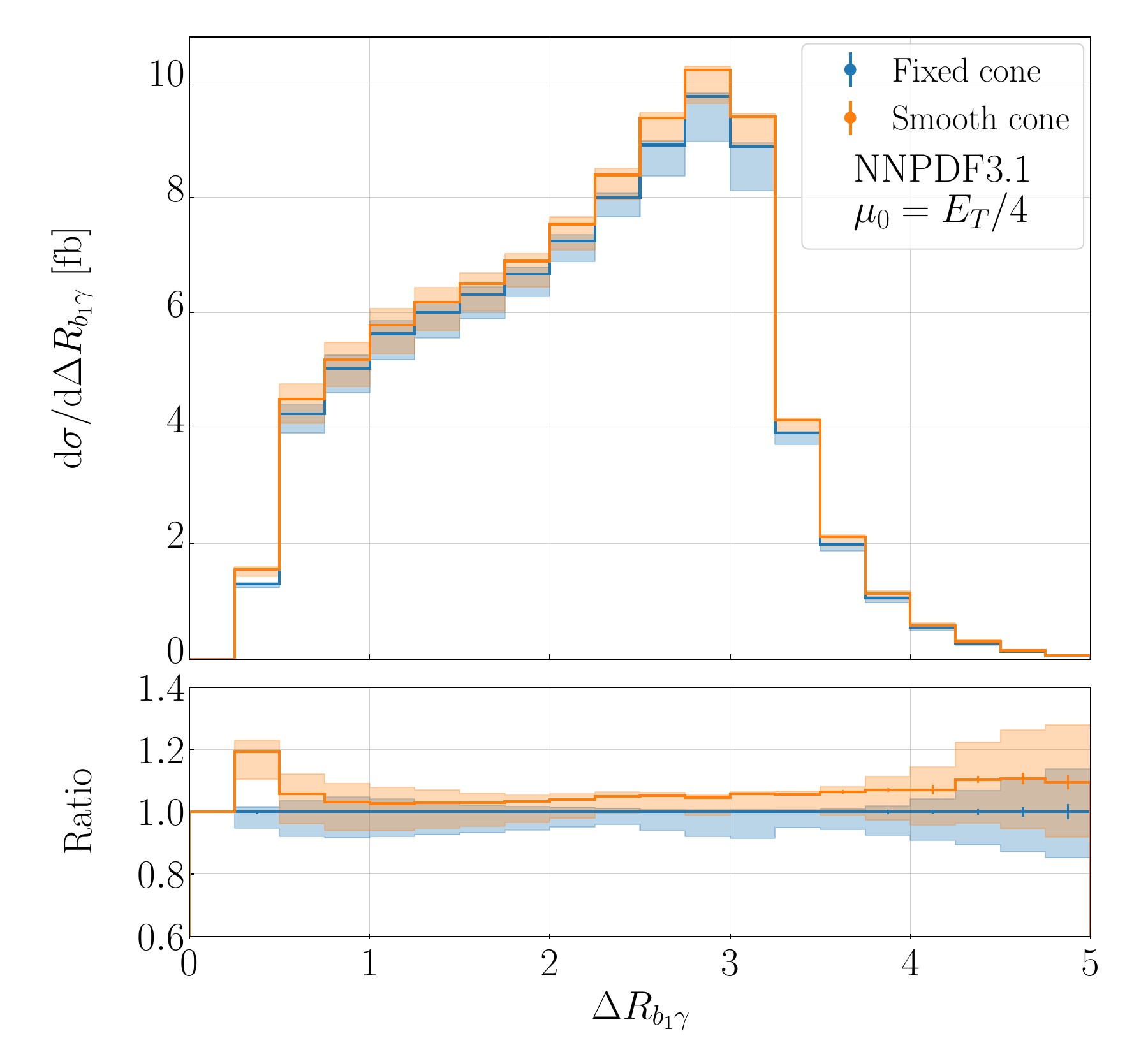}
    \end{center}
    \caption{\label{fig:tta_frag_phot_iso_smooth} \it Differential cross-section distributions 
    at NLO in QCD  for the $pp\to e^+\nu_e\,\mu^-\bar{\nu}_{\mu}\,b\bar{b}\,\gamma+X$ process at the LHC with $\sqrt{s}=13.6$ TeV  as a function of  $\cos\theta_{b_1b_2}$, $p_{T,\,\gamma}$, $\Delta R_{\mu^-\gamma}$ and $\Delta R_{b_1\gamma}$.  Results are presented for the fixed-cone isolation and smooth-cone isolation with $(\varepsilon_{\gamma},n)=(1.00,0.5)$. The lower panels show the ratio to the result obtained with the fixed-cone isolation. Also displayed (as bands) are the theoretical uncertainties coming from a $7$-point scale variation  and (as bars) Monte Carlo integration errors. The scale choice is set to  $\mu_0=E_T/4$. The cross sections are evaluated with the NNPDF3.1 NLO PDF set. }
\end{figure}

In general, however, the use of different photon isolation criteria can modify the normalisation and lead to additional shape distortions at the differential cross-section level when the tuning of the relevant parameters is not performed. To illustrate this, in  Figure \ref{fig:tta_frag_phot_iso_smooth} we present our findings for  $\cos\theta_{b_1b_2}$, $p_{T,\,\gamma}$, $\Delta R_{\mu^-\gamma}$ and $\Delta R_{b_1\gamma}$. The upper panels provide the NLO QCD results with the fixed-cone and smooth-cone isolation prescriptions. In the later case the following input parameters are employed $(\varepsilon_{\gamma},n)=(1.00,0.5)$ with $R=0.4$.  This set of parameters led to the largest differences of about $5\%$ at the integrated cross-section level compared to the calculation with the fixed-cone isolation. Finally, the lower panels display the ratio to the result obtained with the fixed-cone isolation prescription. Also for this comparison the theoretical uncertainties coming from a $7$-point scale variation and Monte  Carlo integration errors are given. 

In the case of $\cos\theta_{b_1b_2}$, the two results differ by about $4\%-5\%$, which is caused mainly by the difference in the normalisation. However, larger effects can be found in particular phase-space regions for other differential cross-section distributions. Specifically, for the transverse momentum of the photon $(p_{T,\,\gamma})$, the differences between the two isolation criteria increase towards the tail of the distribution from about $5\%$ up to $10\%$. On the other hand, for the angular separation between the charged lepton and the photon $(\Delta R_{\mu^-\gamma})$
they are again rather constant and at the level of  $5\%$. Only for  $\Delta R_{\mu^-\gamma}>4$ these differences increase to more than $10\%$. However, these particular phase-space regions are affected by a small number of events.  For the angular separation between the hardest $b$-jet and the photon $(\Delta R_{b_1\gamma})$, we find similar differences of about $10\%$ for larger angular separations. In this case, however, the additional phase-space regions with small angular separations are also affected. For $\Delta R_{b_1\gamma}< 1$  we can observe differences between the two NLO QCD predictions up to even $20\%$. Finally, in all four cases presented here, the theoretical predictions calculated using the smooth-cone isolation lie outside of the uncertainty bands of the results obtained with the help of the fixed-cone isolation. We conclude this section by stating that the random choice of the input parameters $(\varepsilon_{\gamma},n)$ in the smooth-cone isolation criterion  
can introduce additional unnecessary uncertainties and 
may affect current and future comparisons with the  LHC data.

%-----------------------------------------------------------
%
\section{NLO QCD predictions with the fixed-cone isolation prescription}
\label{results-fixed-cone}
%
%-----------------------------------------------------------
%
\begin{table*}[t!]    
    \centering
    \renewcommand{\arraystretch}{1.5}
    \begin{tabular}{l@{\hskip 10mm}l@{\hskip 10mm}l@{\hskip 10mm}c}
        \hline\noalign{\smallskip}
        $\mu_0$& $\sigma^{\rm LO}$ [fb] & $\sigma^{\rm NLO}$ [fb] & $\mathcal{K}=\sigma^{\rm NLO}$/$\sigma^{\rm LO}$ \\
        \noalign{\smallskip}\midrule[0.5mm]\noalign{\smallskip}
        $E_T/4$ & $ 17.512(8)^{+30.9\%}_{-22.1\%} $ & $ 21.50(2)^{+1.4\%}_{-5.0\%} $ & $ 1.23 $ \\
        $H_T/4$ & $ 19.409(9)^{+31.9\%}_{-22.6\%} $ & $ 21.38(2)^{+1.4\%}_{-7.5\%} $ & $ 1.10 $ \\
        $m_t$ & $ 15.877(7)^{+30.1\%}_{-21.6\%} $ & $ 21.13(2)^{+1.4\%}_{-6.4\%} $ & $ 1.33 $ \\
        \noalign{\smallskip}\hline\noalign{\smallskip}
    \end{tabular}
    \caption{\label{tab:tta_frag_kfac} \it Integrated cross sections at LO and NLO  in QCD for the $pp\to e^+\nu_e\,\mu^-\bar{\nu}_{\mu}\,b\bar{b}\,\gamma +X$ process at the LHC with $\sqrt{s}=13.6~{\rm TeV}$. Results are calculated with the fixed-cone isolation using the ALEPH LO quark-to-photon fragmentation function. They are presented for the three scale choices $\mu_0=E_T/4$, $\mu_0=H_T/4$ and $\mu_0=m_t$ with the NNPDF3.1 NLO PDF set. Also displayed are the theoretical uncertainties coming from a $7$-point scale variation ($\pm$ percentages) and Monte Carlo integration errors (in brackets). In the last column the ${\cal K}$-factor is shown. }
\end{table*}
\begin{figure}[t!]
    \begin{center}
	\includegraphics[width=0.49\textwidth]{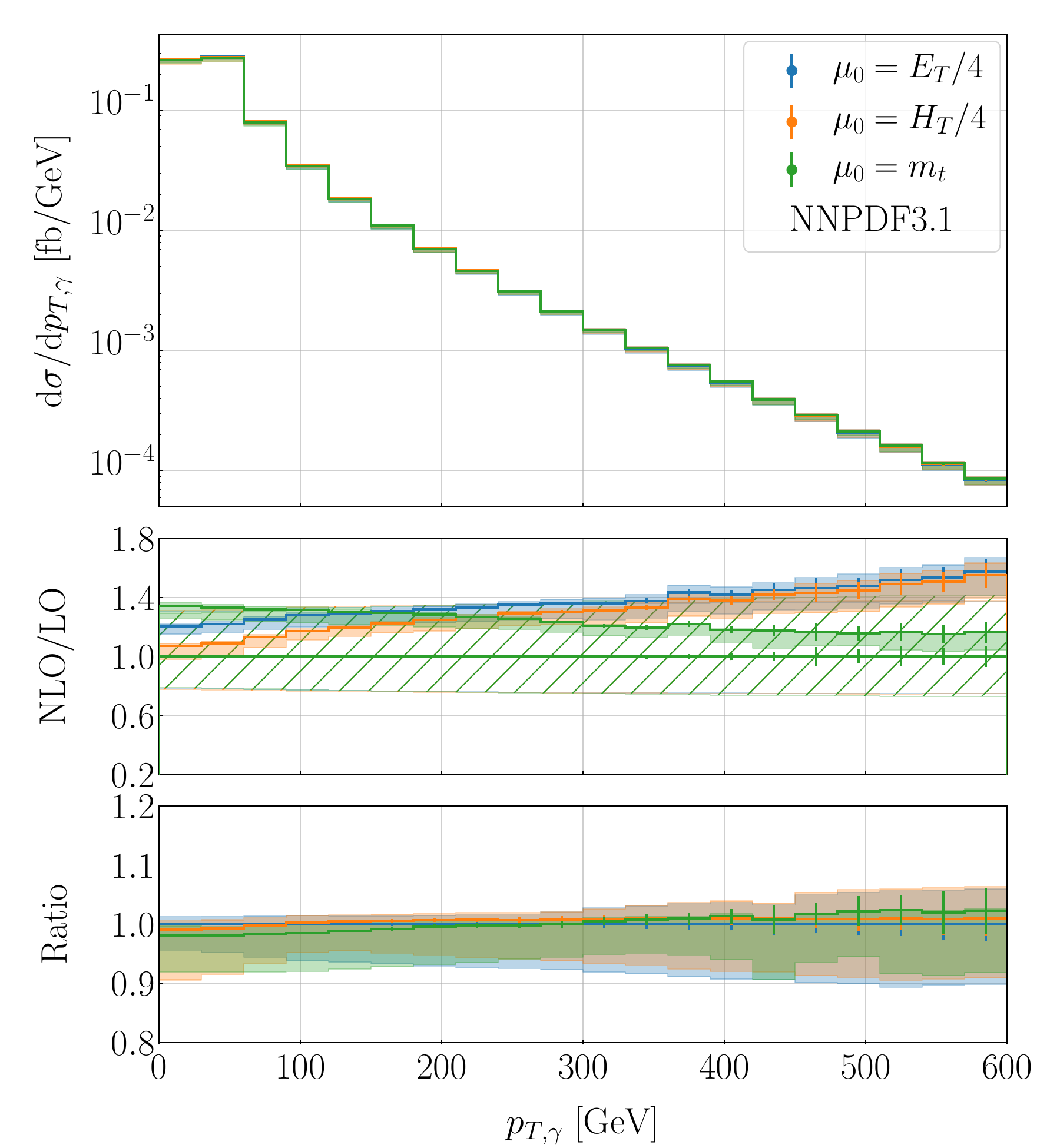}
	\includegraphics[width=0.49\textwidth]{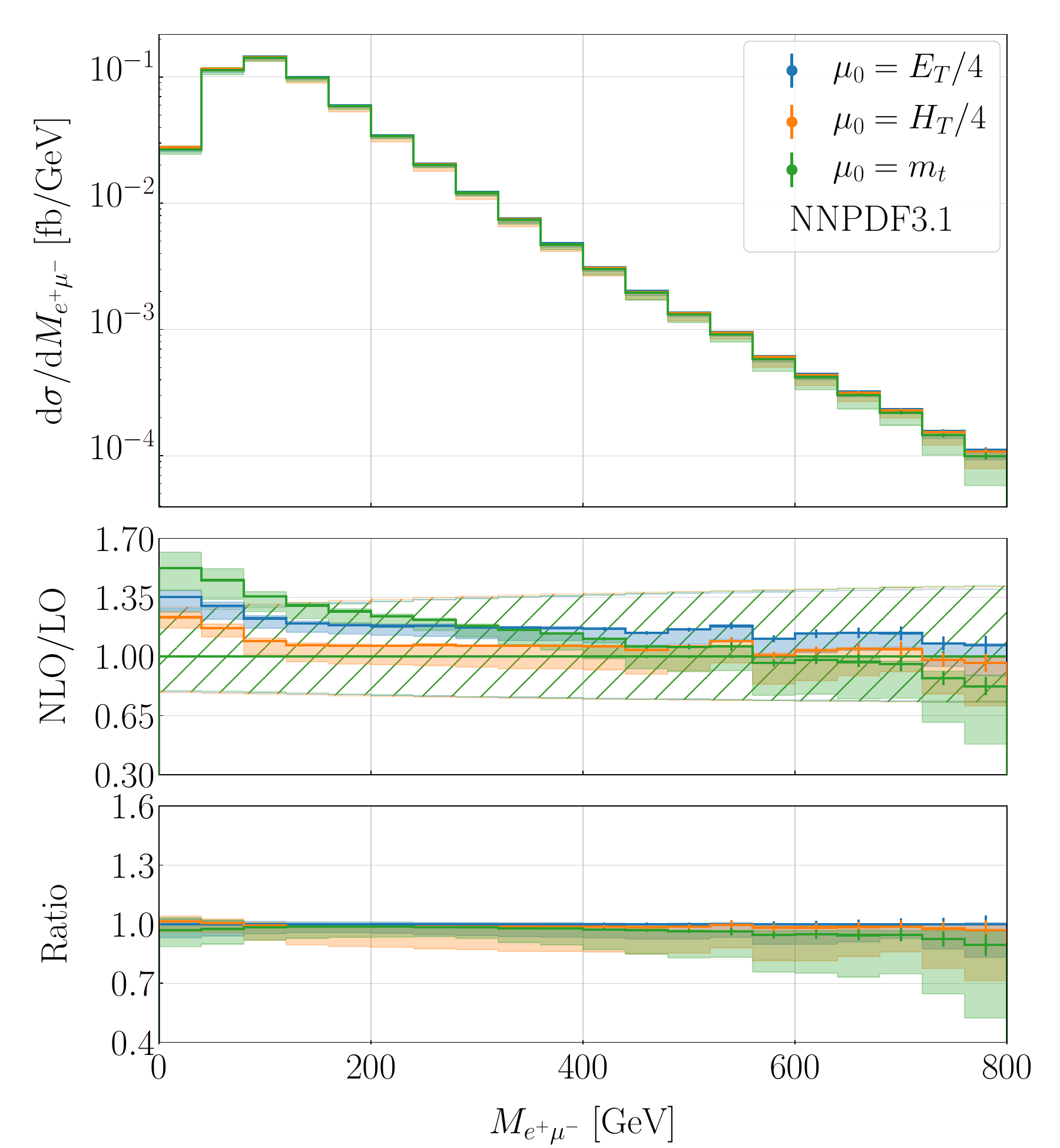}
	\includegraphics[width=0.49\textwidth]{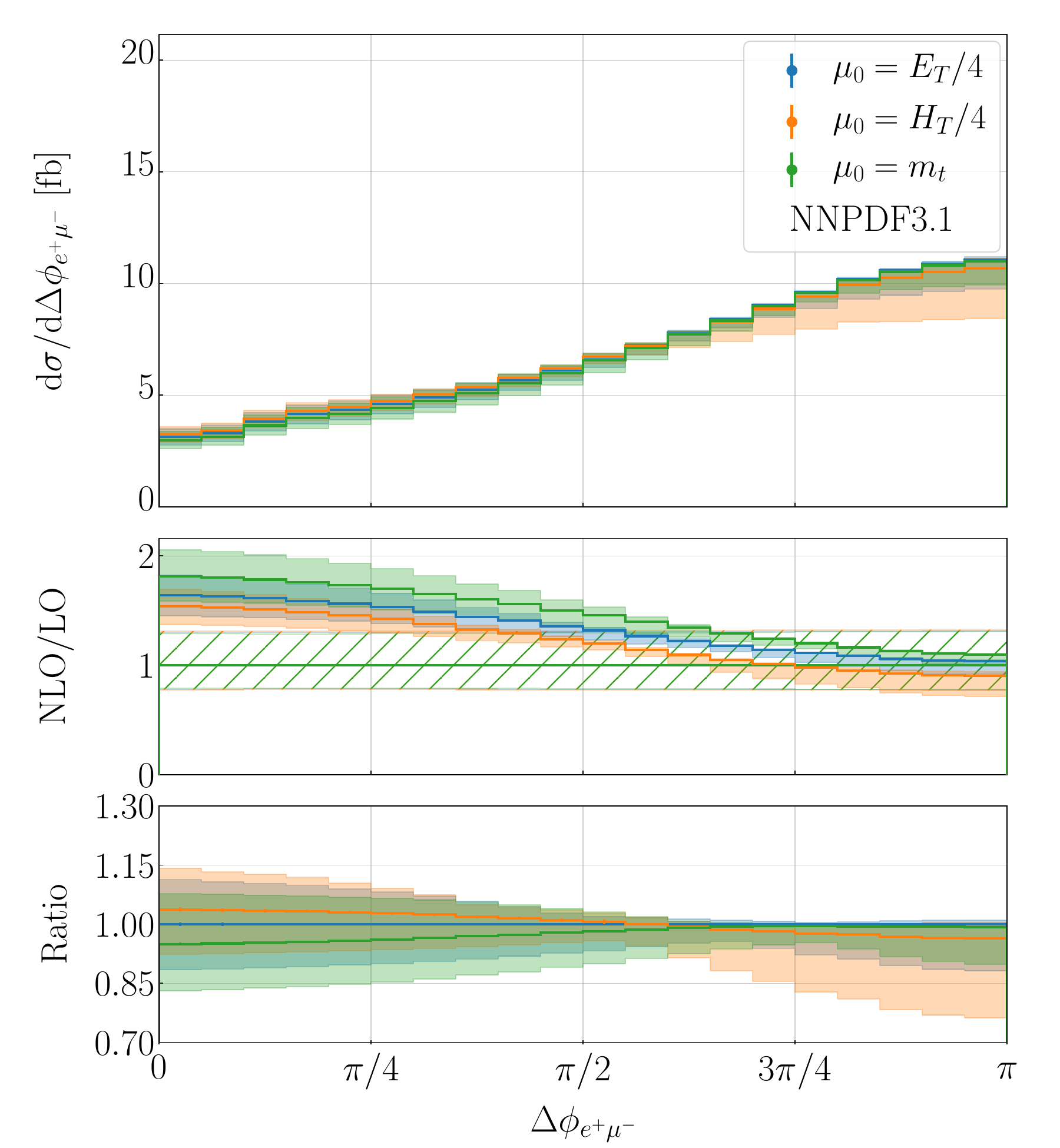}
	\includegraphics[width=0.49\textwidth]{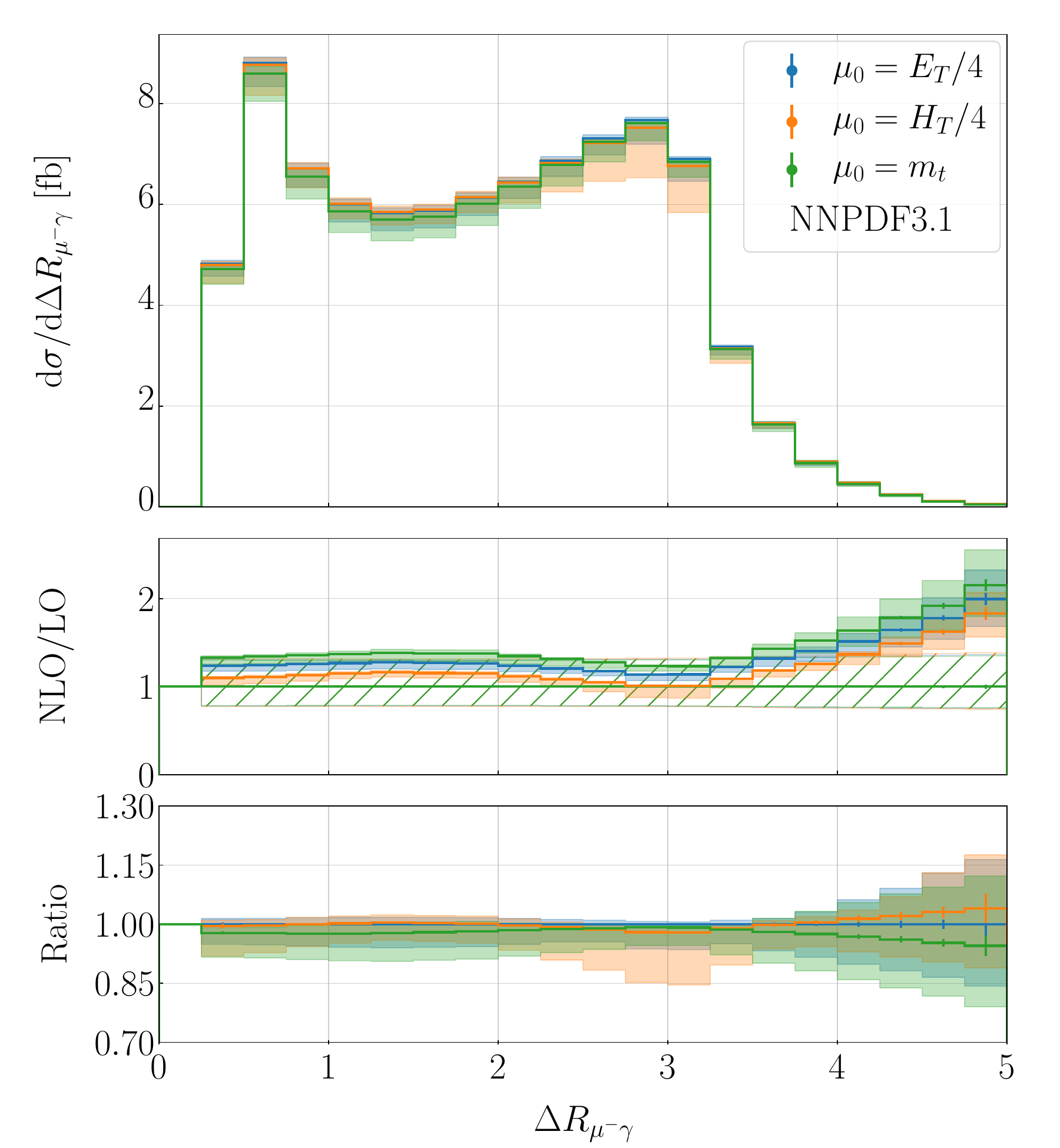}
    \end{center}
    \caption{\label{fig:tta_frag_kfac} \it Differential  cross-section distributions at NLO in QCD for the $pp\to e^+\nu_e\,\mu^-\bar{\nu}_{\mu}\,b\bar{b}\,\gamma +X$ process at the LHC with $\sqrt{s}=13.6$ TeV as a function of $p_{T,\gamma}$, $M_{e^+\mu^-}$,  $\Delta \phi_{e^+\mu^-}$ and $\Delta R_{\mu^-\gamma}$. Results are obtained with the fixed-cone isolation using the ALEPH LO quark-to-photon fragmentation function. They are presented for the three scale settings $\mu_0=E_T/4$, $\mu_0=H_T/4$ and $\mu_0=m_t$ with the NNPDF3.1 NLO PDF set. The middle panels present the differential ${\cal K}$-factor together with its uncertainty band from a $7$-point scale variation and the relative scale uncertainties of the LO cross section. Monte Carlo integration errors are displayed (as bars) in both panels. The lower panels provide the ratio  to the result calculated for  $\mu_0=E_T/4$ together with the corresponding scale uncertainties. }
\end{figure}

In the last part of the paper we present the state-of-the-art theoretical predictions for the $pp\to e^+\nu_e\,\mu^-\bar{\nu}_{\mu}\,b\bar{b}\,\gamma+X$ process with the fixed-cone isolation taking into account  full off-shell effects in the calculations. In our analysis we focus on the NLO QCD results obtained with various scale choices, presented for the LHC III energy of $\sqrt{s}=13.6$ TeV. In particular, we study three different settings covering our default scale choice, $\mu_R=\mu_F=\mu_0=E_T/4$, and the two additional scale settings $\mu_0=H_T/4$ defined in Eq. \eqref{ttaa:scale_mt} and $\mu_0=m_t$. Both scale choices have already been used in previous higher-order calculations for the $pp\to t\bar{t}\gamma$ process in the di-lepton  decay channel \cite{Bevilacqua:2018woc,Bevilacqua:2019quz} and therefore serve as alternative scale settings also in the current study. In Table \ref{tab:tta_frag_kfac} we present the LO and NLO QCD integrated  cross sections together with their corresponding theoretical uncertainties estimated by a $7$-point scale variation. In the  last column we display the ${\cal K}$-factor. The NLO QCD predictions are obtained with the default setup consisting of the two fixed-cone isolation criteria as given in Eq.~\eqref{isolation1} and Eq.~\eqref{isolation2} with the ALEPH LO quark-to-photon fragmentation function. First, we can observe that the NLO QCD corrections are in the range of ${\cal K}=(1.10-1.33)$. The different relative sizes of the NLO QCD corrections are due to the large differences in the LO predictions. Still, all the LO results are within the corresponding LO scale uncertainties that are of the order of $30\%$. On the other hand, at NLO in QCD the differences among the results calculated with the three scale settings are reduced to less than $2\%$. Again they are covered by the corresponding NLO scale uncertainties, that are  of  the order of $5\%$ for $\mu_0=E_T/4$, $8\%$ for $\mu_0=H_T/4$ and $6\%$ for $\mu_0=m_t$.

In Figure \ref{fig:tta_frag_kfac} we present the following differential cross-section distributions: the transverse momentum of the photon $(p_{T, \,\gamma})$, the invariant mass of the two charged leptons $(M_{e^+\mu^-})$,  the azimuthal-angle difference  between these two charged leptons $(\Delta \phi_{e^+\mu^-})$  and the angular separation in the $\phi-y$ plane between the muon and the photon $(\Delta R_{\mu^-\gamma})$. The upper panels show the absolute NLO QCD predictions and the corresponding scale uncertainties. The middle panels display the differential ${\cal K}$ factors together with their uncertainty bands and the relative scale uncertainties of the LO cross sections. The lower panels provide the ratio to the NLO result calculated with $\mu_0=E_T/4$. 

In the case of  $p_{T,\,\gamma}$, we find that the behaviour of the NLO QCD corrections is fairly different for the three scale choices. In particular, higher-order QCD corrections decrease from $34\%$ in the bulk of the distribution to $15\%$ in the tail for $\mu_0=m_t$. On the other hand, the NLO QCD corrections grow steadily to about $55\%$ towards the tails for the two dynamical scale settings exceeding the LO uncertainty bands in these phase-space regions. At the same time, the magnitude of the higher-order effects is different at the beginning of the spectrum for these two scale settings. We obtain NLO QCD corrections of the order of $10\%$ for $\mu_0=H_T/4$ and $20\%$ for $\mu_0=E_T/4$. Overall, the three NLO results differ by less than $2\%$, which is negligible compared to their corresponding scale uncertainties, which are $10\%$ for $\mu_0=E_T/4$, $8\%$ for $\mu_0=H_T/4$ and $6\%$ for $\mu_0=m_t$. 

For $M_{e^+\mu^-}$  the size of higher-order corrections  for the two dynamical scale settings is quite similar and substantially different from the $\mu_0=m_t$ case. Overall, we observe large NLO QCD corrections at the beginning of the spectrum $(23\%-52\%)$, that decrease  towards the tails. As could be expected, the largest shape distortions, up to even $70\%$, are obtained for  $\mu_0=m_t$.  In this particular case, we indeed can find  ${\cal K}=1.52$ in the first bin and ${\cal K}=0.82$ in the last one we plotted. On the other hand, for the dynamical scale settings these shape distortions are only of the order of $30\%$. We note here that the differences in the differential ${\cal K}$-factors for the three scale choices are mainly due to the very different underlying LO distributions. We can clearly see that the absolute NLO QCD predictions agree well with each other. Furthermore, the largest scale uncertainties, of about $40\%$, are obtained in the tail for $\mu_0=m_t$. They are even larger than the corresponding LO scale uncertainties estimated for these phase-space regions. The NLO scale uncertainties  are reduced to $10\%-15\%$ for $\mu_0=E_T/4$ and $20\%-25\%$ for $\mu_0=H_T/4$.  Finally, the differences between the two results obtained with the dynamical scale choices are less than $3\%$. They increase to $10\%$ when the fixed scale setting is employed instead.

For the dimensionless observables, $\Delta \phi_{e^+\mu^-}$ and $\Delta R_{\mu^-\gamma}$, we can observe substantial differences in the differential $\mathcal{K}$-factors for $\mu_0=m_t$, $\mu_0=E_T/4$ and $\mu_0=H_T/4$, that are again driven by the underlying LO results. The largest higher-order QCD corrections, up to $80\%-90\%$, occur for $\mu_0=m_t$. They are significantly reduced for the other two dynamical scale settings.  At the same time, the differences between the NLO results for the three selected scales are at the level of only a few percent. The NLO scale uncertainties are very similar for $\mu_0=E_T/4$ and $\mu_0=m_t$ and are in the range of $5\%-15\%$. For $\Delta R_{\mu^-\gamma}$ they increase to about $20\%$ for large angular separations. On the other hand, for $\mu_0=H_T/4$, the scale uncertainties become larger in certain phase-space regions, such as $\Delta \phi_{e^+\mu^-}> \pi/2$ and  when $\Delta R_{\mu^-\gamma}\approx 3$. Looking at the middle panels, we can see that in some bins the NLO QCD results are outside the LO uncertainty bands.

We conclude that our default scale choice, $\mu_0=E_T/4$, is preferred over the other two scale settings. On the one hand, a dynamic scale setting is generally necessary for more accurate predictions of differential cross-section distributions, especially in high-energy tails. On the other hand, our alternative dynamical scale choice, $\mu_0=H_T/4$, leads to larger scale uncertainties, which can be even twice as large as those obtained for $\mu_0=E_T/4$ or $\mu_0=m_t$. We note, however, that in the high-energy tails the predictions obtained with the two dynamical scales become very similar.  Finally, the fixed scale setting can be used as an alternative scale choice for various angular distributions.

%-----------------------------------------------------------
%
\section{Summary}
\label{summary}
%
%-----------------------------------------------------------
%

In this paper we have presented the first NLO QCD predictions for the $pp\to e^+\nu_e\,\mu^-\bar{\nu}_{\mu}\,b\bar{b}\,\gamma+X$ process with full off-shell effects, which do not require the use of  the smooth-cone isolation prescription. Instead, we have employed the fixed-cone isolation that is used by both the ATLAS and CMS collaborations at the LHC for processes with prompt photons. This isolation prescription allows contributions from collinear photon radiation off QCD partons and requires the inclusion of parton-to-photon fragmention processes. Incorporating the ALEPH LO and BFGII photon fragmention functions into the \textsc{Helac-Dipoles} program has allowed us to compare the fixed-cone, smooth-cone and hybrid-photon isolation criteria within the same Monte Carlo framework. We have quantified the impact of different photon-isolation prescriptions on the integrated and differential cross sections for the LHC Run III energy of $\sqrt{s}=13.6$ TeV. In the first step we have showed that it is possible to achieve a very good agreement between the integrated (fiducial) cross-section results obtained with the fixed-cone and smooth-cone isolation prescriptions if  in the latter case the appropriate tuning of the $(\varepsilon_\gamma,n)$ input parameters is performed. However, such tuning, which  depends on the process and decay channel under consideration as well as the phase-space cuts used, is rather impractical and  also time-consuming. On the other hand, using arbitrary values for the  $(\varepsilon_\gamma,n)$ parameters can lead to differences between the two isolation criteria that are similar in size to the NLO QCD scale uncertainties  and thus are part of the systematic error for this process. We have shown that instead of relying solely on the smooth-cone isolation, it is also possible to use the hybrid-photon isolation. Indeed, we have confirmed that the dependence on the input parameters of the smooth-cone isolation could be greatly reduced, well below $0.5\%$, for the hybrid-photon isolation prescription. In each case we have studied, good agreement with the calculations obtained for the fixed-cone insolation has been observed without any tuning. 

We have reached qualitatively similar conclusions for the differential cross-section distributions that we have examined. However, in this case, the use of different photon isolation criteria not only changed the overall normalisation but also introduced different shape distortions in specific regions of the phase space, if the tuning has not been performed. In addition, also here the random choice of input parameters in the smooth-cone isolation criterion could introduce unnecessary theoretical uncertainties that might affect current and future comparisons with the LHC data.

Finally, we have provided the state-of-the-art theoretical predictions for the $pp\to e^+\nu_e\,\mu^-\bar{\nu}_{\mu}\,b\bar{b}\,\gamma+X$ process with the fixed-cone isolation and taking  into account 
all resonant and non-resonant Feynman diagrams, interferences, and finite-width effects of  the top quarks and $W^\pm/Z$ gauge bosons. We have calculated the NLO QCD corrections to the integrated and differential cross sections for the following three scale settings: $\mu_0=E_T/4$, $\mu_0=H_T/4$ and $\mu_0=m_t$. Depending on the scale setting the full $pp$ cross section has received positive and small to moderate NLO QCD corrections. Specifically, the following range of the  ${\cal K}$-factors  have been obtained ${\cal K}=(1.10-1.33)$. The differences in the ${\cal K}$-factor are mostly due to the large spread 
in the LO predictions. In addition, including higher-order effects has reduced the theoretical error from approximately $30\%-32\%$ to $5\%-8\%$ depending on the scale setting. We have also examined a few  differential cross-section distributions and concluded that our default scale choice, $\mu_0=E_T/4$, is preferred over the other two scale settings. Generally, a dynamic scale setting is  necessary for more accurate predictions of dimensionful differential cross-section distributions, especially in high-energy tails. However, our alternative dynamical scale setting, $\mu_0=H_T/4$, resulted in larger scale uncertainties in some specific corners of the phase space. Therefore, it is always recommended to carry out a dedicated study involving the selection of several scale settings for the actually used fiducial phase-space regions. On the other hand, the fixed scale setting, $\mu_0=m_t$, can be safely used as an alternative scale choice for various angular cross-section distributions. We also note that with the current higher-order calculations for the $pp\to e^+\nu_e\,\mu^-\bar{\nu}_{\mu}\,b\bar{b}\,\gamma+X$ process, future comparisons with the
LHC data will become more accurate and precise.

\acknowledgments{
This work was supported by the Deutsche Forschungsgemeinschaft (DFG) under grant 396021762 - TRR 257: \textit{Particle Physics Phenomenology after the Higgs Discovery}.

Support by a grant of the Bundesministerium f\"ur Bildung und Forschung (BMBF) is additionally acknowledged.

The authors gratefully acknowledge the computing time provided to them at the NHR Center NHR4CES at RWTH Aachen University (project number \texttt{p0020216}). This is funded by the Federal Ministry of Education and Research, and the state governments participating on the basis of the resolutions of the GWK for national high performance computing at universities. }

%\bibliography{references} 

\bibliographystyle{JHEP}

\providecommand{\href}[2]{#2}\begingroup\raggedright\endgroup

\end{document}